\documentclass[aps,prx,twocolumn,superscriptaddress,showkeys,showpacs]{revtex4-2}
\usepackage[utf8]{inputenc}
\usepackage[T1]{fontenc}
\usepackage{multirow}
\usepackage{etoolbox} 
\usepackage{lipsum} 
\usepackage{bm}
\usepackage{hyperref}
\usepackage{makeidx}
\usepackage{amsmath}
\usepackage{amsfonts}
\usepackage{amssymb}
\usepackage{graphicx}
\usepackage{dcolumn}
\usepackage{color}
\usepackage{graphics}
\usepackage{epsfig}
\usepackage{float}
\usepackage[english]{babel}
\usepackage{epstopdf}
\usepackage{graphics, subfigure}
\usepackage{array}
\usepackage{bbold}
\usepackage{natbib}
\usepackage{physics}
\usepackage{verbatim}
\usepackage{xcolor}
\usepackage{dsfont}

\global\long\def\bra#1{\left\langle #1 \right|}%

\global\long\def\kket#1{\ket{\ket{#1}}}%

\global\long\def\braket#1#2{\left\langle #1\right. \left| #2 \right\rangle }%

\global\long\def\bs#1{  \boldsymbol{#1}  }%

\global\long\def\tr{\text{tr}}%

\global\long\def\Tr{\text{Tr}}%

\global\long\def\pd{\partial}%

\global\long\def\abs#1{\left|#1\right|}%

\newcounter{pr}

\newcounter{sr}

\newcounter{stc}

\begin{document}

\title{
Quantum and Semi-Classical Signatures of Dissipative Chaos in the Steady State
}

\author{M. A. R. \surname{Griffith}}
\email{griffithrufo@csrc.ac.cn}
\affiliation{Beijing Computational Science Research Center, Beijing 100193, China}
\address{CeFEMA-LaPMET, Physics Department, Instituto Superior T\'{e}cnico, Universidade de Lisboa Av. Rovisco
Pais, 1049-001 Lisboa, Portugal}
\address{UFAM, Universiade Federal do Amazonas Av. General Rodrigo Octavio Jordão Ramos, 1200 - Coroado I, Brazil}

\author{S. \surname{Rufo}}
\email{srufo@csrc.ac.cn}
\affiliation{Beijing Computational Science Research Center, Beijing 100193, China}
\address{CeFEMA-LaPMET, Physics Department, Instituto Superior T\'{e}cnico, Universidade de Lisboa Av. Rovisco
Pais, 1049-001 Lisboa, Portugal}

\author{Stefano Chesi}
\email{stefano.chesi@csrc.ac.cn}
\address{Beijing Computational Science Research Center, Beijing 100193, China}
\affiliation{Department of Physics, Beijing Normal University, Beijing 100875, China}

\author{Pedro Ribeiro}
\email{pedrojgribeiro@tecnico.ulisboa.pt}
\address{CeFEMA-LaPMET, Physics Department, Instituto Superior T\'{e}cnico, Universidade de Lisboa Av. Rovisco
Pais, 1049-001 Lisboa, Portugal}
\affiliation{Beijing Computational Science Research Center, Beijing 100193, China}

\date{\today }

\begin{abstract}
We investigate the quantum–classical correspondence in open quantum many-body systems using the $SU(3)$ Bose-Hubbard trimer as a minimal model. Combining exact diagonalization with semiclassical Langevin dynamics, we establish a direct connection between classical trajectories\ –\ characterized by fixed-point attractors, limit cycles, or chaos\ –\ and the spectral and structural properties of the quantum steady state.
We show that classical dynamical behavior, as quantified by the sign of the Lyapunov exponent, governs the level statistics of the steady-state density matrix: non-positive exponents associated with regular dynamics yield Poissonian statistics, while positive exponents arising from chaotic dynamics lead to Wigner-Dyson statistics. Strong symmetries constrain the system to lower-dimensional manifolds, suppressing chaos and enforcing localization, while weak symmetries preserve the global structure of the phase space and allow chaotic behavior to persist.
To characterize phase-space localization, we introduce the phase-space inverse participation ratio (IPR$_\phi$), which defines an effective dimension $D$ of the Husimi distribution's support. We find that the entropy scales as $S \propto \ln N^D$, consistently capturing the classical nature of the underlying dynamics. This semiclassical framework, based on stochastic mixtures of coherent states, successfully reproduces not only observable averages but also finer features such as spectral correlations and localization properties.
Our results demonstrate that dissipative quantum chaos is imprinted in the steady-state density matrix, much like in closed systems, and that the interplay between dynamical regimes and symmetry constraints can be systematically probed using spectral and phase-space diagnostics. These tools offer a robust foundation for studying ergodicity, localization, and non-equilibrium phases of open quantum systems.
\end{abstract}

\maketitle

\section{Introduction}
\label{sec:Intro}

Dissipative interactions in quantum systems are typically associated with the loss of coherence and the eventual settling into trivial, classical-like steady states. Yet, in strongly nonlinear systems, coupling to an environment can engender a rich variety of dynamical regimes – including periodic oscillations and even chaos. Understanding how such complex behavior arises and how it can be characterized in open quantum systems is a central challenge in nonequilibrium physics, with implications across quantum optics \citep{gardiner2000quantum}, transport \citep{schaller2014openquantum}, quantum thermodynamics \citep{binder2018thermodynamics}, quantum control \citep{wiseman2014quantum}, and information processing \citep{nielsen2010quantum}. Across these areas, an especially relevant regime of open-system dynamics is the Markovian evolution governed by a Lindblad master equation $\frac{d\rho}{dt} = \mathcal{L}[\rho]$ \citep{breuer2002thetheory,gardiner2000quantum}. Here, $\rho$ is the system’s density matrix and $\mathcal{L}$ is the Liouvillian superoperator, which in standard form can be written as
\begin{equation}
\mathcal{L}[\rho] =-i[H,\rho]+\sum_{l}\left(W_{l}\rho W_{l}^{\dagger}-\frac{1}{2}W_{l}W_{l}^{\dagger}\rho-\frac{1}{2}\rho W_{l}W_{l}^{\dagger}\right),
\end{equation}
accounting for unitary evolution under the Hamiltonian $H$ and for dissipation via a set of memoryless environment-induced dissipative channels, modeled by the action of jump operators $\left\{ W_{l}\right\} $ acting on the system’s degrees of freedom. This Lindblad framework, despite its simplicity, has enabled substantial progress in the theoretical description of open quantum systems. However, a fundamental open question remains: How do signatures of complex or chaotic dynamics manifest in the steady state and long-time behavior of an open quantum system, and how are they related to the system’s underlying classical dynamics? 

In this work, we aim to address this question by examining the imprint of dissipative quantum chaos on the steady state and long-time  semiclassical dynamics. As an illustrative case study, we consider a three-site Bose–Hubbard model (trimer) with dissipation – a model that admits chaos in the classical limit without requiring external driving~\cite{PhysRevA.108.013314} and, as we will discuss, allows us to develop a systematic treatment of quantum fluctuations on top of the classical evolution. Most importantly, we show how the semiclassical features of the long-time dynamics are directly linked to level statistics and other structural properties of the quantum steady state.  

The general structure of the paper is as follows: 
The reminder of Section \ref{sec:Intro} discusses previous work and highlights our main results. 
Section \ref{sec:ModelandMethods} introduces the Bose–Hubbard trimer model and the methods used to analyze it. Section \ref{sec:Results} presents our results. Section \ref{sec:Discussion} provides further discussion, systematization, and interpretation of the results. Finally, Section \ref{sec:Conclusion} offers our conclusions.

\begin{figure}
\centering \includegraphics[width=0.9\linewidth]{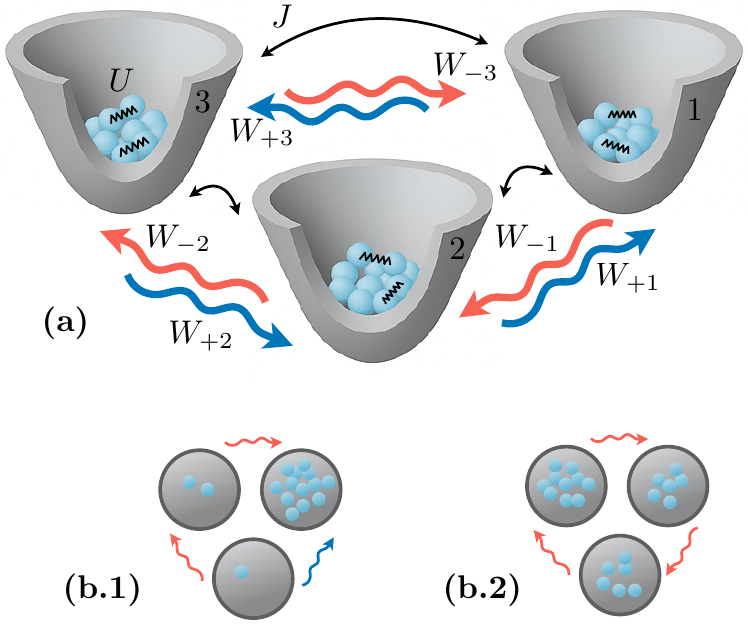}
\caption{(a) Illustration of the Bose-Hubbard trimer model of Eq.~(\ref{Eq:Hamiltonian})
in a presence of coherent ($J$) and incoherent ($W_{\pm i}$) inter-site hopping. 
$U$ is the strength of the intra-well bosonic interactions. 
The two limiting cases in the bottom panels refer to different dissipative regimes: (b.1) favors accumulation of bosons into the first well, yielding a single steady state, and (b.2) induces a cyclic movement of bosons, yielding a non-stationary classical orbit.
}
\label{fig:Model} 
\end{figure}

\subsection{Previous Work} 
The past decade has witnessed substantial progress in understanding chaos and integrability in open quantum systems. On one front, significant advances have been made in simulating and solving Lindblad dynamics. In one dimension, for example, nonequilibrium extensions of matrix product state methods were implemented to study driven-dissipative quantum chains \citep{zwolak2004mixedstate,verstraete2004matrixproduct}, and tensor-network techniques have been developed for higher-dimensional open systems \citep{kshetrimayum2017asimple}. At the same time, several classes of exactly solvable (integrable) Lindblad superoperators have been identified. These include certain quadratic (non-interacting) bosonic and fermionic Liouvillians \citep{prosen2008thirdquantization,prosen2010quantization,barthel2022solving}, interacting spin and fermion models with engineered dissipation \citep{prosen2011exactnonequilibrium,prosen2014exactnonequilibrium,ilievski2017dissipationdriven,landi2022nonequilibrium,ziolkowska2020yangbaxter}, and fully connected or central-spin models \citep{rowlands2018noisyspins,ribeiro2019integrable,claeys2022dissipative,arXiv.2407}. 
Parallel efforts have classified Lindbladians into symmetry-based universality classes \citep{lieu2020tenfold,altland2021symmetry}. On another front, there has been growing interest in defining dissipative quantum chaos by analogy with the well-known signatures of chaos in closed Hamiltonian systems. In particular, a quantum-chaos conjecture for open systems has been proposed \citep{denisov2019universal,can2019randomlindblad,can2019spectral,sa2020spectral,sa2020complex} in direct analogy to the Bohigas-Giannoni-Schmit conjecture for closed systems \citep{berry1977levelclustering,bohigas1984characterization,casati1980onthe}. It posits that the spectral statistics of the Liouvillian superoperator reflects the integrability of the dynamics: integrable (exactly solvable) Liouvillians exhibit Poissonian level-spacing statistics, whereas nonintegrable (chaotic) Liouvillians exhibit level repulsion following random-matrix (Wigner-Dyson) statistics. A convenient measure for this purpose is the distribution of complex spacing ratios in the Liouvillian spectrum \citep{sa2020complex}, which has been used as an empirical tool to distinguish “regular” vs. “chaotic” open dynamics. These predictions have been supported by extensive numerical data on random or generic Liouvillians \citep{sa2020complex,sa2020spectral2,sa2021integrable,rubio-garcia2022fromintegrability}. Other diagnostic tools like the spectral form factor, widely used in closed-system quantum chaos, have also been generalized~\citep{xu2021thermofield,li2021spectral,haake2018quantum}. Recent progress also includes establishing an analog of  the eigenstate thermalization hypothesis (ETH) for Liouvillians \cite{Almeida2025, Ferrari2025}

Open quantum systems exhibit unique features with no direct analog in closed systems. One important distinction is the notion of steady-state integrability versus full Liouvillian integrability. For instance, a system may admit an analytic solution for its nonequilibrium steady state (indicating a form of integrability in the stationary properties) even if the entire Liouvillian spectrum is nonintegrable. A notable example is the boundary-driven XXZ spin chain studied by Prosen and \v{Z}nidari\v{c}~\citep{prosen2013eigenvalue}, which possesses a closed-form steady state yet is believed to be nonintegrable as a Liouvillian (indeed, its spectrum shows signatures of chaos \citep{sa2020complex}). Another distinction is that chaotic characteristics in the Liouvillian spectrum do not always manifest in the system’s long-time dynamics. In generic open systems with a unique attractive steady state, any initial condition eventually relaxes exponentially to a stationary state, obscuring direct dynamical signatures of spectral chaos. Recent work has shown, for example, that the separation between regular and chaotic Liouvillian spectra may break down in the long-time behavior of certain dissipative systems, effectively rendering their asymptotic dynamics classical-like regardless of spectral signatures~\cite{2024_PRL_David}. On the other hand, some non-generic open systems can evade convergence to a stationary state by entering oscillatory persistent phases (sometimes termed time crystals) that never decay to a single steady state~\cite{iemini2018boundary, arXiv.2411.07297}. 

Many of the differences between closed and open quantum chaos can be understood from a semiclassical perspective. In a Hamiltonian system, classical chaos provides an accurate description of quantum dynamics only up to the Ehrenfest time, after which quantum interference becomes dominant and the infinite-time limit does not commute with the classical limit. By contrast, in an open system, environmental noise and dissipation damp out quantum coherence, so one may expect the long-time limit and the classical limit to commute \cite{kamenev2011fieldtheory,PhysRevLett.128.070402,arXiv.2407.20314}. Indeed, numerical studies have found that introducing dissipation can restore classically predicted behavior that is absent in the closed quantum case. For example, in the paradigmatic kicked rotor, quantum localization effects (absence of energy diffusion) are destroyed by coupling to a bath, recovering the classical diffusive energy growth \citep{dittrich1987quantum,dittrich1988effects,dittrich1990longtime,carlo2005dissipative}. More generally, detailed comparisons of classical and quantum dynamics in dissipative systems show good agreement in the statistical properties of chaotic trajectories \citep{carlo2005dissipative,carlo2017classical,carlo2017classical2,carlo2017signatures,carlo2019effects,carlo2019threedimensional}. From the purely classical viewpoint, nonlinear dissipative systems can exhibit a rich spectrum of asymptotic behaviors\ –\ including single or multiple fixed-point attractors, limit cycles, and strange (chaotic) attractors\ –\ depending on model parameters \citep{morrison2008dynamical,kessler2012dissipative,hannukainen2018dissipationdriven,ferreira2019lipkinmeshkovglick,kilin1978effectof,drummond1978volterra,iemini2018boundary,ivanchenko2017classical,stitely2020nonlinear,kolovsky1993steadystate,volokitin2017computation,hartmann2017asymptotic,carlo2017classical,carlo2017signatures,carlo2017classical2,carlo2019threedimensional}.  Transitions between regular and chaotic regimes (for example, via period-doubling cascades \citep{ivanchenko2017classical,carlo2017classical,carlo2017signatures,volokitin2017computation,carlo2017classical2,hartmann2017asymptotic}) are often accompanied by qualitative changes in the quantum steady state of the corresponding system\ –\ effectively, nonequilibrium phase transitions between different dynamical phases. However, the precise quantum fingerprints of these different classical regimes, such as how the steady-state level statistics or entropic measures change when a classical attractor becomes chaotic, remain not fully understood. 

The study of chaotic versus regular features in quantum steady states was initiated by the pioneering work of Prosen and \v{Z}nidari\v{c}~\citep{prosen2013eigenvalue}. They established that the integrable or nonintegrable nature of the steady state $\rho_0$ in certain dissipative spin chains is reflected in the statistical distribution (Poissonian versus Wigner-Dyson) of an effective Hamiltonian $H_{\text{eff}}$, defined by $\rho_{0}\propto e^{-H_{\text{eff}}}$. Despite this progress, a direct semiclassical understanding of the origin of these statistical signatures\ –\ in analogy with the classical-quantum correspondence well established in closed-system chaos\ –\ is still lacking. The existence of an analogous correspondence in dissipative systems admitting a well-defined classical limit is supported by recent results of Mondal \emph{et al.}~\cite{Mondal2025}, which uncover distinct quantum signatures of dissipative chaos in the steady state of an open Tavis-Cummins dimer.
However, findings by Ferrari \emph{et al.}~\cite{Ferrari2023} reveal that steady-state quantum chaos can occur even when classical or mean-field analyses indicate regular behavior, underscoring the subtle quantum-classical interplay in dissipative dynamics.  Another study by Richter \emph{et al.}~\cite{Richter2025} examined boundary-driven spin chains, showing that a system’s nonequilibrium steady state can be integrable even when the Liouvillian is chaotic (and vice-versa). Such findings highlight the need for a systematic exploration of how classical chaos and quantum steady-state properties relate within a paradigmatic model.

In the present work, we bridge this quantum-classical gap by systematically comparing quantum and semiclassical indicators of chaos in a minimal open model. Specifically, we investigate the $N$-boson dissipative Bose-Hubbard trimer (three-mode cavity or triple-well BEC) as a testbed for quantum chaos in an open many-body system. This model features an $SU(3)$ symmetry in the Hamiltonian and, crucially, can exhibit chaotic dynamics in the appropriate classical (mean-field) limit without any external periodic driving~\cite{PhysRevA.108.013314}. The semiclassical limit is obtained by letting the total particle number $N\to\infty$, which coincides with a mean-field description using $SU(3)$ coherent states. For any finite $N$, the Hilbert space is finite-dimensional (no truncation is needed for numerics) and, except for trivial symmetry-protected cases \citep{lidar2003decoherencefree}, the Lindblad evolution yields a unique steady state $\rho_{0}$. Chaotic behavior in Hamiltonian collective $SU(3)$-symmetric systems was first explored by Meredith \emph{et al.}~\cite{Meredith1988}, with further developments in the context of Bose–Hubbard trimers and related models~\cite{Gnutzmann1999,PhysRevA.108.013314}. Experimentally, proposals for realizing such dynamics have emerged in platforms such as trapped ions~\cite{Jason2012}, where an $SU(3)$-symmetric three-level Lipkin-Meshkov-Glick model can interpolate between integrable and chaotic regimes, and ultracold atoms in optical lattices~\cite{Nemoto2000,Mossmann2006,Franzosi2002,Viscondi_2011,Chianca2011}, where the Bose–Hubbard trimer serves as a natural implementation.

\begin{figure}
\centering \includegraphics[width=1.0\linewidth]{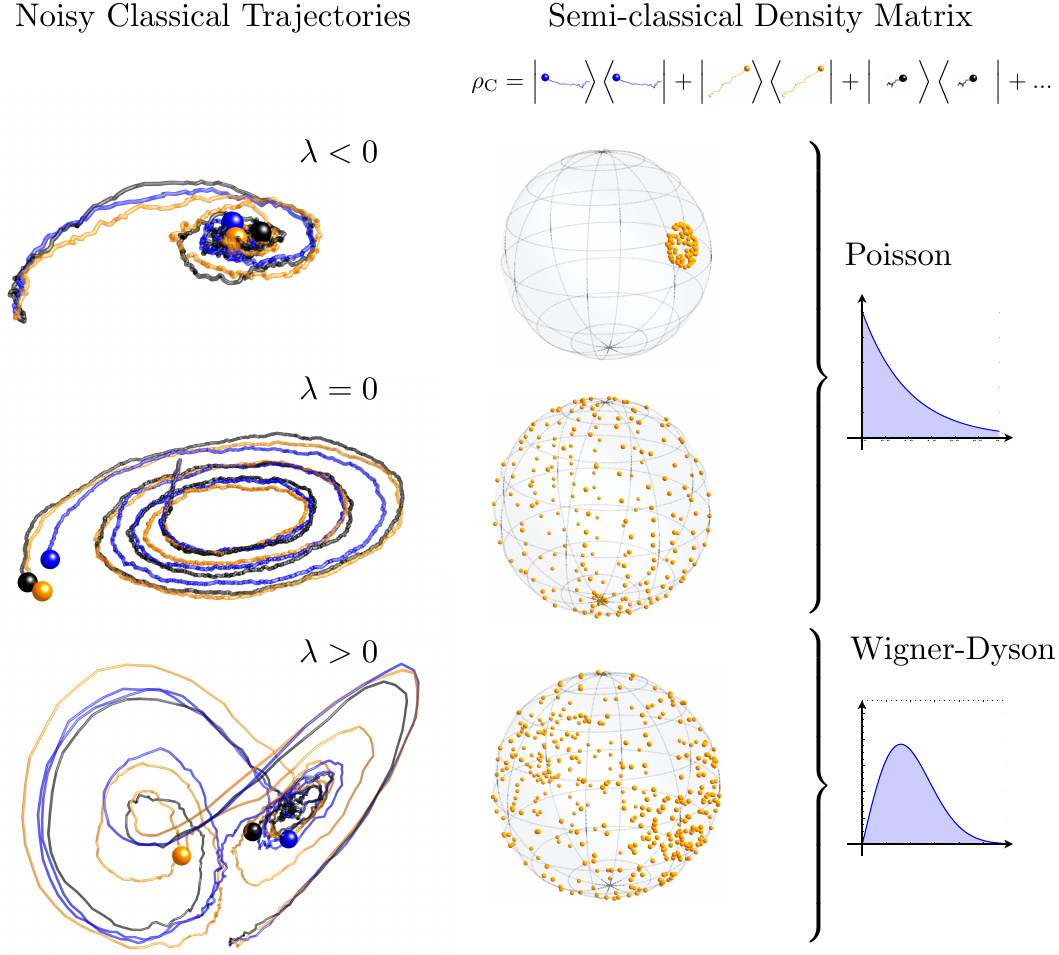}
\caption{ 
Illustration of how noisy classical trajectories\ –\ regular or chaotic\ –\ lead to Poisson or Wigner-Dyson level statistics of the steady state density matrix. Classical trajectories governed by attractor dynamics ($\lambda<0$), regular motion ($\lambda=0$), or chaotic dynamics ($\lambda>0$) are evolved under stochastic Langevin dynamics, generating a classical ensemble of coherent states. This ensemble defines a semiclassical density matrix through long-time phase-space distributions. For regular dynamics, trajectories concentrate near attractors or explore smooth regions of phase space, resulting in localized Husimi distributions and Poissonian statistics. In contrast, chaotic dynamics yields rough and delocalized distributions across phase space, giving rise to spectral correlations described by Wigner-Dyson statistics.
}
\label{fig:summ.results} 
\end{figure}

\subsection{Main Results}
We identify two qualitatively distinct dissipative regimes in the long-time dynamics of the classical Bose-Hubbard trimer, as illustrated in Fig.~\ref{fig:Model}. The first is an \emph{attractor configuration}, in which dissipation drives the system toward a single (or multiple) stable fixed-point attractor(s). The second is a \emph{cyclic configuration}, where dissipation leads to sustained competing population cycles among the three modes, resulting in persistent limit-cycle behavior. These classical regimes correspond, in the quantum setting, to fundamentally different spectral structures of the Liouvillian: The attractor configuration yields a \emph{gapped} Liouvillian spectrum, with an isolated zero eigenvalue associated with the unique steady state; the cyclic configuration produces a \emph{gapless} spectrum, in which the separation between the steady state and subleading eigenvalues vanishes with increasing system size. 

We study the steady state and the Liouvillian spectrum by performing exact diagonalization and time evolution for the quantum system, alongside a semiclassical analysis of the mean-field equations augmented by Langevin stochastic noise of order $1/\sqrt{N}$ \cite{kamenev2011fieldtheory}. 
To probe the onset and signatures of dissipative chaos in the quantum system, we employ a range of diagnostic tools. These include the level-spacing statistics of an effective Hamiltonian $H_{\text{eff}}$ (defined via $\rho_{0} \propto e^{-H_{\text{eff}}}$), the von Neumann entropy of the steady state, and a phase-space inverse participation ratio (IPR$_\phi$) derived from the Husimi representation of $\rho_{0}$, which quantifies the degree of localization in phase space. On the classical side, we compute the Lyapunov exponents and analyze Poincar\'e sections of the semiclassical trajectories, to identify regular versus chaotic motion. To account for quantum fluctuations, we add perturbative $1/N$ corrections to the classical dynamics via a stochastic Langevin term. Averaging the resulting trajectories as mixtures of coherent states defines a ``classical density matrix'' \(\rho_\mathrm{C}\). This construction allows us to investigate the correspondence between classical chaotic dynamics and quantum signatures encoded in the steady state.

Our results establish a direct correspondence between the nature of the classical long-time dynamics and both spectral properties and structural features of the quantum steady state.

--- \emph{First}, when the classical system evolves toward an isolated fixed-point attractor (characterized by a negative Lyapunov exponent and absence of chaos), the quantum steady state remains regular: the Liouvillian retains a finite spectral gap in the large-$N$ limit, and the level statistics of the effective Hamiltonian $H_{\text{eff}}$ follow a Poisson distribution --  hallmarks of integrable dynamics.

--- \emph{Second}, in regimes where the classical dynamics becomes non-stationary --  due to the emergence of limit cycles or strange (chaotic) attractors --  the Liouvillian gap closes as $N \to \infty$. In these cases, the level statistics of $H_{\text{eff}}$ transitions to Wigner-Dyson (random matrix theory) behavior \emph{only when} the classical dynamics is chaotic, i.e., exhibits a positive Lyapunov exponent. In contrast, purely regular (non-chaotic) orbits, charactrized by a non-positive Lyapunov exponent, still lead to Poissonian statistics. This firmly links the presence of quantum chaotic signatures in the steady state to classical chaos in the underlying mean-field dynamics.

--- \emph{Third}, steady states associated with classically chaotic dynamics exhibit structural features consistent with quantum ergodicity: They are highly delocalized in phase space (as seen from a small inverse participation ratio) and possess a large von Neumann entropy $S(\rho_{0})$. In contrast, steady states arising from regular dynamics remain localized in phase space and show lower entropy, reflecting a higher degree of purity and organization. In intermediate regimes --  where classical phase space supports both regular and chaotic trajectories --  we observe mixed level statistics in $H_{\text{eff}}$, interpolating between Poisson and Wigner-Dyson distributions, consistent with partial ergodicity in the quantum steady state.

--- \emph{Fourth}, the classical mixture $\rho_\text{C}$ not only captures observable averages to next-to-leading order in $1/N$, but also reproduces finer features of the steady state, including level statistics, entropy, and phase-space delocalization. This indicates that, for large $N$, Lindblad steady states --  even those associated with limit cycles or chaotic attractors --  can be effectively understood within a classical framework.

Taken together, our results demonstrate a robust quantum-classical correspondence in open quantum systems: The spectral and structural features of the quantum steady state faithfully reflect the nature of the long-time classical dynamics. Specifically, while classical chaos is marked by a closing Liouvillian gap and the emergence of Wigner-Dyson statistics --  indicative of quantum chaoticity -- integrable classical behavior results in Poissonian level statistics in the effective Hamiltonian $H_{\text{eff}}$. At the same time, the presence of an attractor or regular classical orbits correspond to a Liouvillian which is respectively gapped or gapless. The degree of phase-space delocalization and the entropy of the steady state further reinforce the quantum-classical correspondence, distinguishing between ergodic and non-ergodic regimes.

Our findings provide a semiclassical explanation for the integrability-chaos dichotomy in steady-state level statistics, as originally observed by Prosen and \v{Z}nidari\v{c}~\citep{prosen2013eigenvalue} in one-dimensional chains with no classical counterpart. Moreover, they clarify an important subtlety in the broader spectral characterization of open quantum systems: while Liouvillians exhibiting regular spectral statistics -- as diagnosed by the complex spacing ratio~\citep{sa2020complex} -- consistently yield steady states with Poissonian $H_{\text{eff}}$ statistics, Liouvillians conforming to Ginibre (GinUE) spectral statistics can give rise to steady states that are regular, chaotic, or mixed in terms of their internal eigenvalue structure. This highlights the crucial distinction between the spectral properties of the full Liouvillian and those of the effective Hamiltonian encoding the steady-state structure, and helps clarifying which signatures of dissipative quantum chaos are ultimately encoded in the steady state itself.

Figure~\ref{fig:summ.results} provides a schematic summary of our central findings, establishing a direct correspondence between classical and quantum features. This figure encapsulates the emergence of quantum chaotic signatures from underlying classical chaos, and sets the stage for the detailed numerical and analytical analysis presented in the remainder of the paper.

\section{Model and Methods}
\label{sec:ModelandMethods}

We consider a system of $N$ bosons distributed among three wells,
as illustrated in Fig.~\ref{fig:Model}, and described by the Bose-Hubbard
Hamiltonian: 
\begin{equation}
\mathcal{H}=-\sum_{i=1}^{3}J_{i}\left(b_{i}^{\dagger}b_{i+1}+b_{i+1}^{\dagger}b_{i}\right)+\frac{U}{2N}\sum_{i=1}^{3}b_{i}^{\dagger}b_{i}\left(b_{i}^{\dagger}b_{i}-1\right),\label{Eq:Hamiltonian}
\end{equation}
where $U$ is the on-site interaction and $J_{i}$ are the hopping amplitudes
(with $b_{4}\equiv b_{1}$). In the following, we usually consider equal amplitudes and
normalize the energies to $J_{i}=J$, thus setting $J=1$. We also assume that, besides
coherent tunneling, bosons can move between wells incoherently. This process can be induced, e.g., by noise in the hopping coupling constant
\cite{ 
PhysRevLett.123.080601,
SciPostPhys.6.4.045} and is modeled by the jump
operators 
\begin{equation}
W_{-i}=\sqrt{w_{-i}}\,b_{i+1}^{\dagger}b_{i},\,\,\,\,W_{+i}=\sqrt{w_{+i}}\,b_{i}^{\dagger}b_{i+1},\label{eq:jump}
\end{equation}
where the index $-(+)$ denotes a clockwise (anti-clockwise) process
and $w_{\pm i}$ are jumping rates, see Fig.~\ref{fig:Model} for a
pictorial representation. This incoherent dynamics still conserves the
total number of particles $N=\sum_{i}b_{i}^{\dagger}b_{i}$. In the
$N$-particle sector, the Hilbert space has
dimension $d=\frac{1}{2}(N+1)(N+2)$, being spanned by the states $|n_{1},n_{2},n_{3}=N-n_{1}-n_{2}\rangle$, where
$0\le n_{i}\le N$ is the occupation number of mode $i$. Within each sector, the action
of the bilinear operators $b_{i}^{\dagger}b_{j}$ is isomorphic to $SU(3)$,
see Appendix~\ref{App-A-SU3-Gen} for the explicit mapping. 

In the following, it is convenient to consider a vectorized notation
for the density matrix $\rho\to\ket{\ket{\rho}}=\sum_{i,j}\rho_{i,j}\ket i\bra j^{T}$,
in which the Liouvillian writes 
\begin{align}
\mathcal{L} = & -i\left[H\otimes\mathds{1}-\mathds{1}\otimes H^{T}\right]+\frac{1}{N}\sum_{\mu,i}\left[W_{\mu i}\otimes W_{\mu i}^{\dagger T}\right.\nonumber \\
  & \left.-\frac{1}{2}(W_{\mu i}^{\dagger}W_{\mu i})\otimes\mathds{1}-\frac{1}{2}\mathds{1}\otimes(W_{\mu i}^{\dagger}W_{\mu i})^{T}\right].\label{eq:L_vector}
\end{align}
The Liouvillian spectrum $\left\{ \Lambda_{0}=0,\Lambda_{1},\Lambda_{2}...\right\} $,
with $\Lambda_{0}$ corresponding to a steady state $\mathcal{L}\kket{\rho_{0}}=0$,
is defined such that $\text{Re}\Lambda_{i}\ge\text{Re}\Lambda_{j}$
if $i<j$ (note that $\text{Re}\Lambda_{i} \leq 0$). We are interested in the properties of the asymptotic long-time
regime of a generic Liouvillian for which the steady-state is unique, thus $\lim_{t\to\infty}\rho(t)=\rho_{0}$. 

\begin{figure}[th!]
\centering \includegraphics[width=8.3cm]
{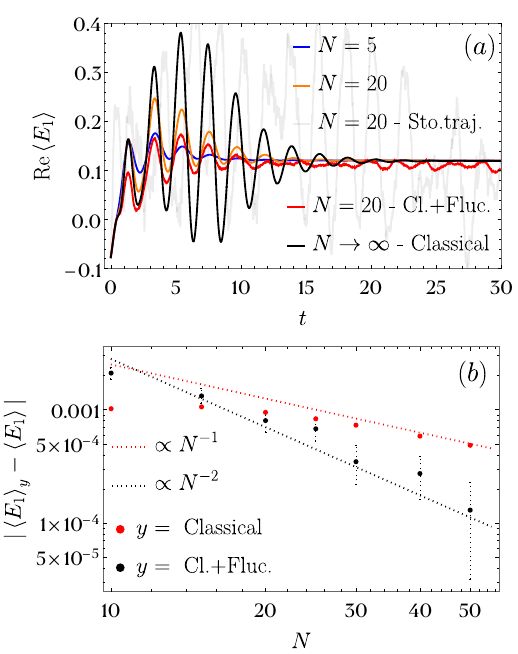}
\caption{(a) Evolution of the real part of the expectation value $ \left\langle  E_1 \right\rangle = \langle b_3^\dagger b_2 \rangle $, plotted for different values of $N$. With $N=20$ we  also show the stochastic dynamics,  averaged over $1000$ trajectories (red), and a typical single trajectory (gray). (b) Scaling with $N$ of the difference between $\left\langle  E_1 \right\rangle$ and the classical approximations $\left\langle  E_1 \right\rangle_y$, where $y$ indicates presence ($y= \text{Cl.+Fluc.} $) or absence ($y= \text{Classical} $) of quantum fluctuations. Dotted lines are guides to the eye.  }
\label{fig:QuantClass} 
\end{figure}

In its vectorized form, the spectrum and eigenstates of $\mathcal{L}$
can be obtained by exact diagonalization techniques. However, as the
dimension of the matrix $\mathcal{L}$ scales with the square of the
Hilbert space dimension, i.e. $d^{2}\propto N^{4}$, this already
presents a considerable computational challenge for $N\gtrsim20$.
In the following, we present exact diagonalization results of the
full eigensystem for $N$ up to 15. For larger numbers of bosons,
$N=20,30$, we take advantage of the sparse structure of $\mathcal{L}$,
applying the Lanczos algorithm to obtain the Liouvillian gap, and
rely on sparse matrix multiplication techniques to numerically evolve a generic
initial state until a suitable convergence to the steady state is
attained. 

\subsection{Semi-classical Dynamics}

For large $N$, the bosonic system is amenable to a semi-classical
treatment that can be systematically performed by considering $SU(3)$
coherent states 
\begin{equation}
\ket{\boldsymbol{\eta}}=e^{b_{2}^{\dagger}b_{1}\eta_{2}+b_{3}^{\dagger}b_{1}\eta_{3}}\ket{N,0,0},
\end{equation}
with $\boldsymbol{\eta}=(\eta_{2},\eta_{3})^{T}$ and $\eta_{j=2,3}\in\mathbb{C}$ These coherent states form an over-complete basis \citep{Perelomov1986}, see also Appendix~\ref{App-B-SU3-CS}. 

In the presence of dissipation, the propagator for the density matrix
can be expressed in a path-integral form on the Keldysh contour \cite{Sieberer_2016, kamenev2011fieldtheory}. We provide the explicit derivation of this procedure using $SU(3)$ coherent states in Appendix~\ref{App-C-SC-DissD}.
By extremising the resulting action, we obtain the classical equations
of motion 
\begin{eqnarray}\label{evolution_classical}
\partial_{t} \boldsymbol{\eta} = \boldsymbol{X(\eta)}, \label{semiclassical}
\end{eqnarray}
where 
\begin{eqnarray} \label{eq:X}
 & \boldsymbol{X(\eta)} & =-i\boldsymbol{\omega}^{-1}\boldsymbol{\partial_{\bar{\eta}}}\mathcal{H}\label{semiclassical_X}\\
 &  & +\frac{1}{2}\sum_{\mu,i}\mathcal{\bar{W}}_{\mu i}\left(\boldsymbol{\omega}^{-1}\boldsymbol{\partial_{\bar{\eta}}}\mathcal{W}_{\mu i}\right)-\left(\boldsymbol{\omega}^{-1}\boldsymbol{\partial_{\bar{\eta}}}\mathcal{\bar{W}}_{\mu i}\right)\mathcal{W}_{\mu i},\nonumber 
\end{eqnarray}
with $\boldsymbol{\partial_{\bar{\eta}}}=(\partial_{\bar{\eta}_{2}},\partial_{\bar{\eta}_{3}})^{T}$ where $\bar{\eta}_{i}=\eta_{i}^{*}$. $\mathcal{H}$ and $\mathcal{W}_{\mu i}$ are, respectively, the average of the Hamiltonian and jump operators in the coherent states basis:
\begin{equation}
\mathcal{H}=\frac{1}{N}\frac{\bra{\boldsymbol{\eta}}H\ket{\boldsymbol{\eta}}}{\braket{\boldsymbol{\eta}}{\boldsymbol{\eta}}},  \ \  \mathcal{W}_{\mu i}=\frac{1}{N}\frac{\bra{\boldsymbol{\eta}}W_{\mu i}\ket{\boldsymbol{\eta}}}{\braket{\boldsymbol{\eta}}{\boldsymbol{\eta}}}.
\end{equation}
Throughout this article we denote complex conjugation with a bar and, in particular, $\bar{\mathcal{W}}_{\mu i} = \mathcal{W}_{\mu i}^*$. Finally, the matrix $\boldsymbol{\omega}$ is defined by $\omega_{j,k}=N^{-1}\partial_{\bar{\eta}_{j}}\partial_{\eta_{k}}\ln\braket{\boldsymbol{\eta}}{\boldsymbol{\eta}}$.

Solutions of Eq.~(\ref{semiclassical}) define trajectories $\boldsymbol{\eta}(t)=(\eta_{2}(t),\eta_{3}(t))\in\mathbb{C}^{2}$. 
For convenience, we represent the two-dimensional complex phase-space using the inverse stereographic projection to the  unit sphere in 5 dimensions, $\mathbb{C}^{2} \to S_{4} \in \mathbb{R}^{5}: \boldsymbol{\eta} \to \boldsymbol{x} $, with  
\begin{eqnarray}
\boldsymbol{x}= & \frac{\left(2\text{Re}\eta_{2},\,2\text{Im}\eta_{2},\,2\text{Re}\eta_{3},\,2\text{Im}\eta_{3},\,|\eta_{2}|^{2}+|\eta_{3}|^{2}-1\right)^{T}}{1+|\eta_{2}|^{2}+|\eta_{3}|^{2}}.\label{eq:inv_stereo}
\end{eqnarray}

Starting with an initial coherent state, the classical trajectory follows the quantum dynamics up to a time of order $1/\sqrt{N}$. Figure~\ref{fig:QuantClass}(a) shows the time evolution for several values of $N$ in a case where the system attains a steady state that is well captured by the classical approximation, $N\to\infty$. In this case, classical and quantum evolutions also agree at large times.

To improve the semi-classical description, we consider the next-to-leading order correction in $1/N$. Within the Keldysh path integral, this can be done by taking into account quantum fluctuations up to quadratic order. This procedure,  derived in Appendix~\ref{App-D-StocCP}, yields an effective stochastic term that corrects the deterministic dynamics of Eq.~(\ref{semiclassical})~\cite{kamenev2011fieldtheory}. The resulting Langevin equation is: 
\begin{eqnarray}
\partial_{t} \boldsymbol{\eta} =  \boldsymbol{X(\eta)} + \frac{1}{\sqrt{N}} \boldsymbol{z},\label{eq:LangevinEq}
\end{eqnarray}
where $\boldsymbol{z}=(z_{2},z_{3})^{T}$ is a complex-valued
white noise variable with expectation value $\mathbb{E}[\bar{z}_{i}(t) z_{j}(t')]= g_{i,j}(\boldsymbol{\eta}) \delta(t-t')$ and $\mathbb{E}[z_{i}(t) z_{j}(t')]= f_{i,j}(\boldsymbol{\eta}) \delta(t-t')$.
It is worth noting that noise is multiplicative, as the coefficient matrices $g(\boldsymbol{\eta})=\bs M_{22}$ and $f(\boldsymbol{\eta})=\bs M_{12}$ (given in Appendix~\ref{app:attract}) depend explicitly  on the phase-space coordinates $\boldsymbol{\eta}$.

For $N=20$, Fig.~\ref{fig:QuantClass}(a) considers the time evolution of a typical observable and compares the exact quantum dynamics with the classical ($N\to\infty$) prediction, as well as with the average over 1000 realizations of the Langevin dynamics. A typical stochastic trajectory is also represented. Including quantum fluctuations via Langevin dynamics substantially improves the classical prediction.

\subsection{Semi-classical Density Matrix}

According to the outlined semi-classical approximation, accounting for quantum fluctuations yields a density matrix that is a classical mixture of coherent states. Explicitly,  $\rho(t)=\rho_{\text{C}}(t)+O(N^{-2})$, with: 
\begin{eqnarray}
\rho_{\text{C}}(t)=\int D\boldsymbol{\xi}P\left[\boldsymbol{\xi}\right]\frac{\ket{\boldsymbol{\eta}_{\left[\boldsymbol{\xi}\right]}(t)}\bra{\boldsymbol{\eta}_{\left[\boldsymbol{\xi}\right]}(t)}}{\braket{\boldsymbol{\eta}_{\left[\boldsymbol{\xi}\right]}(t)}{\boldsymbol{\eta}_{\left[\boldsymbol{\xi}\right]}(t)}},\label{eq:rho_Langevin}
\end{eqnarray}
where $\int D\boldsymbol{\xi}P\left[\boldsymbol{\xi}\right]$ denotes the measure over white-noise and $\boldsymbol{\eta}_{\left[\boldsymbol{\xi}\right]}(t)$ is obtained solving Eq.~\eqref{eq:LangevinEq} with a particular noise realization $\boldsymbol{\xi}(t')$ ($0<t'<t$) and assuming a coherent initial state parametrized by $\boldsymbol{\eta}(0)$. 

In the following, we investigate the properties of the density matrix and of its classical approximation. We evaluate Eq.~\eqref{eq:rho_Langevin} approximately by sampling over noise realizations drawn from the white-noise probability distribution
$P\left[\boldsymbol{\xi}\right]$, 
\begin{eqnarray}
\rho_{\text{C}}(t)\simeq\frac{1}{M_{\text{S}}}\sum_{i=1}^{M_{\text{S}}}\frac{\ket{\boldsymbol{\eta}_{\left[\boldsymbol{\xi}_{i}\right]}(t)}\bra{\boldsymbol{\eta}_{\left[\boldsymbol{\xi}_{i}\right]}(t)}}{\braket{\boldsymbol{\eta}_{\left[\boldsymbol{\xi}_{i}\right]}(t)}{\boldsymbol{\eta}_{\left[\boldsymbol{\xi}_{i}\right]}(t)}},\label{eq:rho_Langevin_sample}
\end{eqnarray}
with $M_{\text{S}}$ the number of samples. 

In Fig.~\ref{fig:QuantClass}(b), we show the finite-size scaling of the error of the mean value of a typical observable computed for large times. Taking into account the fluctuations by averaging over several Langevin noise realizations (black points) is considerably better than considering the noiseless classical trajectory (red points). Although the scaling with $N$ is not fully conclusive for the sizes available to us, it is compatible with the expected $N^{-2}$ and $N^{-1}$  dependences.

\subsection{Spectral properties of the density matrix}

In the following, we characterize the steady-state density matrix, $\rho_{0}$, in terms of its spectrum. 
Specifically, we parametrize $\rho_{0}$ by its effective  Hamiltonian $\rho_0 = e^{-H_\text{eff}}$ and let $\varepsilon_i$ denote the eigenvalues of $H_\text{eff}$, arranged in ascending order. 
One natural quantity of interest is the von Neumann entropy, given by $S = -  \text{Tr}\rho \ln \rho = \sum_i e^{-\varepsilon_i} \varepsilon_i$.  

Another quantity of interest is the level statistics, measured by the level-spacing ratios $r_i = (\varepsilon_{i+1} - \varepsilon_{i})/(\varepsilon_{i} - \varepsilon_{i-1})$ of the eigenvalues. 
In Hamiltonian systems, level statistics is widely used as a heuristic method to distinguish chaotic from regular dynamics, following the conjectures of Berry and Tabor~\cite{Berry1977} and of Bohigas, Giannoni, and Schmit~\cite{Bohigas1984, Casati1980}. Accordingly, quantum counterparts of classically integrable systems follow Poisson statistics, while systems with a chaotic semi-classical limit conform to the level statistics of one of the Wigner-Dyson random-matrix ensembles, depending on their symmetries. Here, we use this approach to analyze $H_\text{eff}$. 

\begin{figure*}
\centering
\includegraphics[width= \textwidth ]{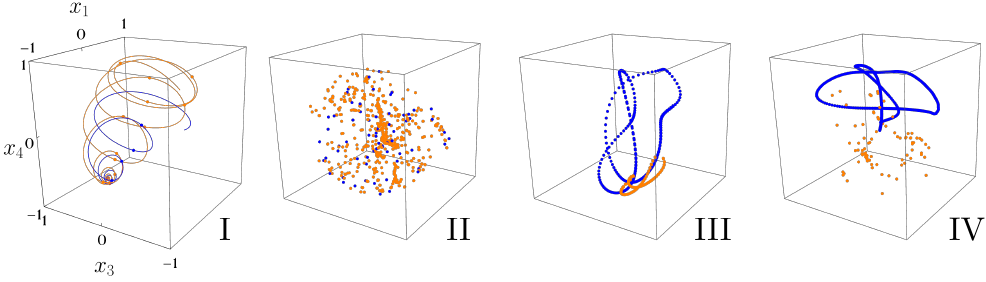} %
\caption{ Poincar\'e cross sections of the classical dynamics projected in the 3-dimensional space $(x_1,x_3,x_4)$.
Points represent the intersections of the trajectories with the $x_2 = 0$ plane for cases I--IV, computed for two different initial conditions (orange and blue). 
In I, the full projected trajectory is also shown (lines). The parmaeters are as chosen as follows. Case I: $U=1,\, w_{+1}= 1.4,\, w_{-2}= 1.1,\, w_{-3}= 1.2$; Case II: $U=6, \,w_{-1}= 1.1,\,  w_{-2}= 1.2,\, w_{-3}= 1.4$; Case III: $U=0.01,\,w_{-1}= 0.1,\,  w_{-2}= 0.5, \, w_{-3}= 0.4$; Case IV: $U=4,\,w_{+1}= 0.2,\, w_{-1}= 0.5, \, w_{-2}=w_{-3}= 0.7$.}
\label{fig:Poincare-CS}
\end{figure*}

\subsection{ Phase-space Inverse Participation Ratio}

The Inverse Participation Ratio (IPR) ~\cite{VisscherIPR} is a powerful tool to quantify localization of a distribution over set of states. The IPR is widely used to study localization-delocalization transitions in single- and many-particle states. Here, to diagnose the degree of localization of chaotic and regular steady states, we propose to generalize the IPR to phase-space according to the following definition, which measures the degree of spreading over the basis of (non-orthogonal) coherent states:  
\begin{eqnarray}\label{eq:IPR_phi}
\text{IPR}_\phi[\rho] =  \frac{\int d\mu(\boldsymbol\eta)  \, \mathbb{H}_\rho(\boldsymbol\eta)^2}{\left[\int d\mu(\boldsymbol\eta) \, \mathbb{H}_\rho(\boldsymbol\eta) \right]^2},
\end{eqnarray}
where $\mathbb{H}$ is the Husimi distribution  of $\rho$, given by $\mathbb{H}_\rho(\boldsymbol\eta) = \bra{\boldsymbol\eta} \rho \ket{\boldsymbol\eta} / \braket{\boldsymbol\eta}{\boldsymbol\eta}$, and $d\mu(\boldsymbol\eta)$ is the invariant measure over the coherent-state manifold, which for $SU(3)$ is given by
\begin{eqnarray}\label{eq:measure_su3}
d\mu(\boldsymbol\eta)  =  \frac{(N+1)(N+2)}{2}  \int 
 \frac{ d\text{Re}\eta_2\, d\text{Im}\eta_2\, d\text{Re}\eta_3 \, d\text{Im}\eta_3 }{ \frac{\pi^2}{2}
\left(1+\abs{\eta_2}^2 + \abs{\eta_3}^2\right)^3 }. \nonumber \\ 
\end{eqnarray}
In the basis diagonal in the boson numbers,  $\text{IPR}_\phi[\rho]$ can be evaluated explicitly in terms of the matrix elements of the density matrix (see Appendix~\ref{app:IPR}).  

For a coherent-state density matrix of $SU(3)$, of the form  $\rho = \ket{\boldsymbol\eta_0}\bra{\boldsymbol\eta_0}/\braket{\boldsymbol\eta_0}{\boldsymbol\eta_0}$,   
we get $\text{IPR}_\phi[\rho] = (N+2)/(2N+2)$, which is of order $N^0$ in the large $N$ limit.
Instead, for a state uniformly spread in phase space, $\rho=  [(N + 1)(N + 2)/2 ]^{-1}   \mathds{1} $, we get $\text{IPR}_\phi[\rho]= [(N + 1)(N + 2)/2 ]^{-1}$, which scales with the inverse of the Hilbert space dimension, i.e., it decreases as $\sim N^{-2}$ when $N$ is large.
 
The physical interpretation of this quantity can be related through $\text{IPR}_\phi  \sim N^{-D/2}$ to the dimension $D$ of the manifold in which the Husimi distribution is supported.
The argument is as follows:  
As the dimension of the Hilbert space grows as $N^2$, and the phase space is 4-dimensional, a coherent state occupies a phase-space volume of order $\ell^4 \propto N^{-2}$, where $\ell$ is a typical linear scale in phase space.  
Thus, to generate a manifold of dimension $D$ we need $\ell^{-D}\propto N^{D/2}$ states. Assuming that all such states have similar probability, $p \sim N^{-D/2} $, we can estimate $\text{IPR}_\phi \sim N^{D/2} \times ( N^{-D/2} )^2 \sim N^{-D/2}$.



\subsection{ Numerical Methods}

We implemented a finite-difference method to solve the  stochasitc evolution described by Eq.~(\ref{eq:LangevinEq}).  To compute the Lyapunov exponents from Eq.~(\ref{evolution_classical}), we used the JuliaDynamics package of Ref.~\cite{Datseris2018}, providing the Lyapuvov spectrum of the dynamical system. To obtain the steady-state density-matrix operator, we have applied the Lindbladian evolution to an initial quantum state, up to times at which the condition $\partial_t \rho_0 = 0 $ is satisfied with sufficient accuracy. The eigenvalues and eigenvectors of the Lindblad superoperator  have been obtained by employing suitable diagonalization algorithms.

\section{Results} 
\label{sec:Results}

\subsection{Classical Dynamics - Illustrative examples} 

\begin{figure}[t!]
\centering
\includegraphics[width=1\linewidth ]
{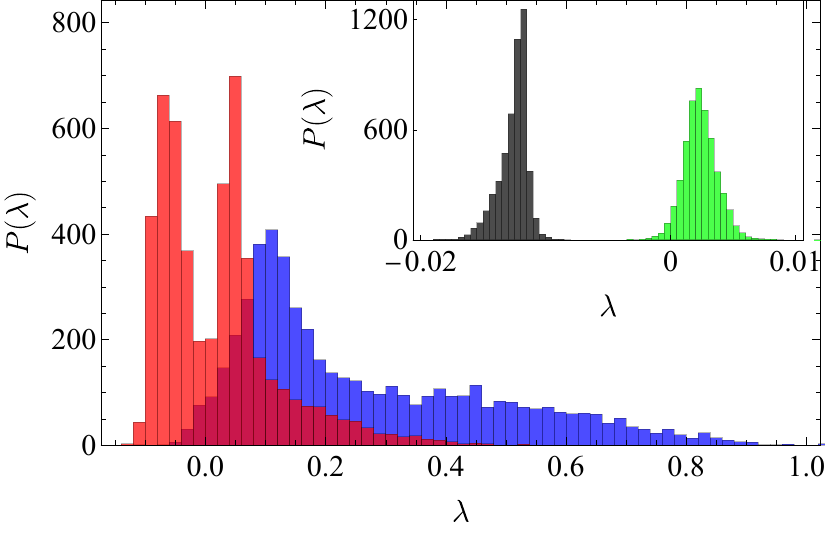} 
\caption{Distribution $P(\lambda)$ of Lyapunov exponents for the same four cases of Fig.~\ref{fig:Poincare-CS}. Here, we use $5000$ initial conditions and $t_f=1000$. The main chart shows the distributions of Cases II (blue) and IV (red), while the inset refers to Cases I (black) and III (green). In the mixed Case~IV, the fraction of positive Lyapunov exponents is about $57 \%$. }
\label{fig:LSSto}
\end{figure}

To illustrate our results, we present four representative cases, by first examining their classical dynamics.  
Figure~\ref{fig:Poincare-CS} shows the inverse stereographic projections [see Eq.\eqref{eq:inv_stereo}] of Poincar\'e cross-sections, obtained by integrating Eq.~\eqref{semiclassical} for sufficiently long times (to overcome the initial transient). A systematic analysis of the chaotic or regular nature of classical trajectories can be done by studying their leading Lyapunov exponent~\cite{datseris2022nonlinear,ChaosBook}, denoted as $\lambda$ (see its formal definition in Sec.~\ref{App-E-Liouv-Gap}). While $\lambda>0$ indicates chaotic dynamics, $\lambda<0$ characterizes attractors, and a vanishing $\lambda$ corresponds to stable limit cycles and periodic trajectories.  
The histograms in Fig.~\ref{fig:LSSto} show the distribution of Lyapunov exponents of classical trajectories with initial states uniformly sampled over the phase space. For Case I, the dissipative terms favor the state where all bosons occupy well~1. As seen in Fig.~\ref{fig:Poincare-CS}(a), the dynamics yields a unique stable steady state, to which all trajectories converge at large times. The distribution $P(\lambda)$, shown in the inset of Fig.~\ref{fig:LSSto} (black histogram), is peaked around a negative value, $\lambda\simeq - 0.01$, which is consistent with the contracting dynamics imprinted by the fixed point. 

For Cases II--IV, the dissipative terms induce a cyclic flow of bosons, and do not favor any particular steady state. Case II has a relatively large interaction strength $U$ and, as seen in Fig.~\ref{fig:Poincare-CS}(b), presents a Poincar\'e map characteristic of fully chaotic dynamics, with no apparent orbits or fixed points. The corresponding $P(\lambda)$, shown by the blue histogram of Fig.~\ref{fig:LSSto}, is mostly supported in the region $\lambda>0$, consistent with strong chaos in almost all phase space.

In contrast, Case III considers a small value of $U$ and, as seen in Fig.~\ref{fig:Poincare-CS}(c), yields regular orbits, with the long-time dynamics approaching limiting cycles.
Concomitantly, the green histogram of Fig.~\ref{fig:LSSto} (inset) is a distribution $P(\lambda)$ which remains very close to $\lambda=0$. Deviations are comparable to the numerical error in determining the Lyapunov exponents, which is consistent with regular dynamics arising throughout all phase space. 

Finally, Case IV [Fig.~\ref{fig:Poincare-CS} (d)] presents mixed features, with parts of phase space showing characteristics of a chaotic attractor, while in others parts regular orbits can clearly be found. The type of long-time dynamics depends on the initial condition. This scenario is confirmed by the red histogram of Fig.~\ref{fig:LSSto}, which shows a $P(\lambda)$ with a double-peak structure. One peak is centered around $\lambda =0 $ and another around a positive value.

%

\begin{figure*}[t!]
\centering
\includegraphics[width=1\linewidth ]{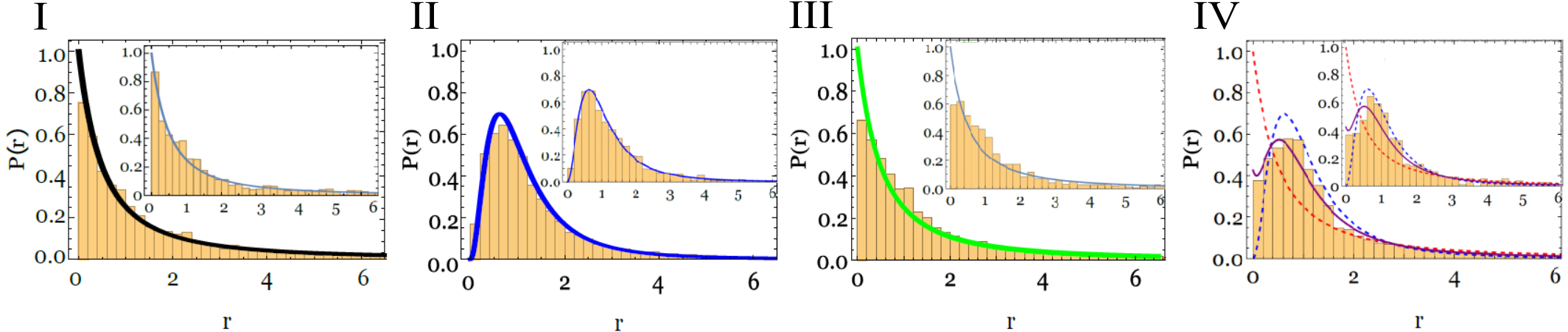} 
\caption{Level-spacing ratio statistics $P(r)$, obtained from the stationary density matrix of representative non-cyclic (I), cyclic (II and III), and mixed (IV) cases. In each panel we show the distribution obtained from the quantum (main) and semiclassical (inset) density matrix, with $N=15$. The solid curves in panels I and III (II) are Poissonian (GUI) distributions. The purple curve of case IV is the mixed distribution $\tilde{P}(r)=(1-\alpha)P_{Poisson}(r) + \alpha P_{GUE}(r)$, with $\alpha = 0.57$. Parameters are as in Fig.~\ref{fig:Poincare-CS}.   }
\label{fig:Avr}
\end{figure*} 


%
\begin{figure}[t!]
\centering
\includegraphics[width=1\linewidth ]{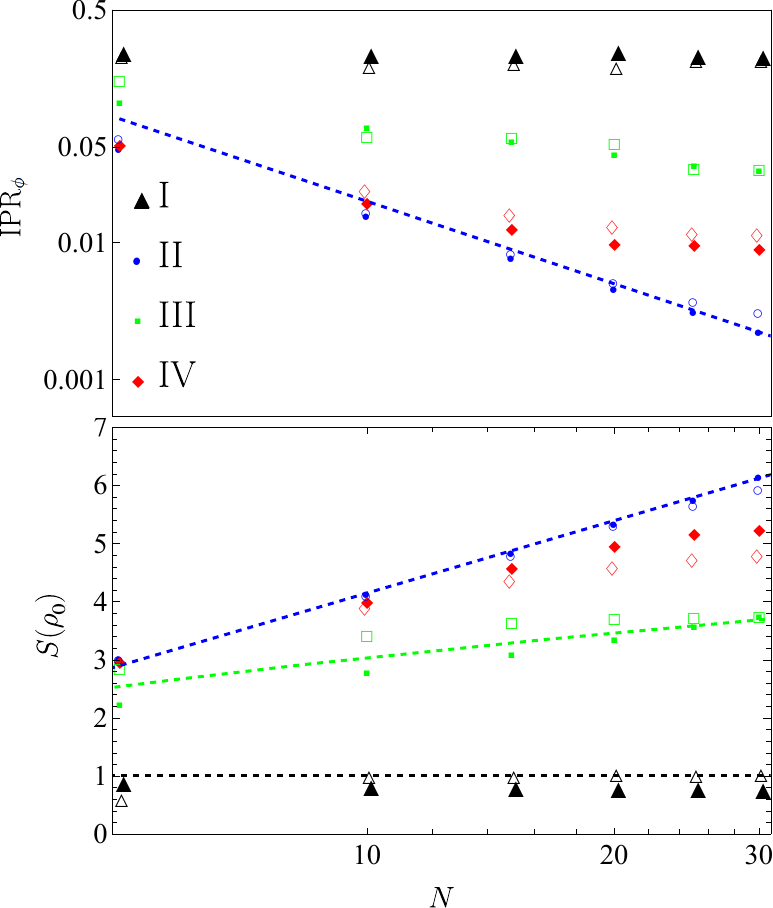}
\caption{Scaling   with $N$ of (a) ${\rm IPR}_\phi$ and (b) von Neumann entropy $S(\rho_0)$. In both panels, Cases I (black triangle), II  (blue circle), III (green square), and IV (red diamond) are the same of Fig.~\ref{fig:Poincare-CS}. The solid symbols are found from the quantum density matrix, while the open symbols refer to the stochastic density matrix $\rho_{C}$. Dashed lines indicate various power-laws. Case I: constant ${\rm IPR}_\phi$ and $S(\rho_0)$; Case II: ${\rm IPR}_\phi \propto N^{-2}$ and $S(\rho_0) \propto \ln N^2$; Case III: ${\rm IPR}_\phi \propto N^{-1}$ and $S(\rho_0) \propto \ln N$.
}
\label{fig:EntropIPR}
\end{figure} 

\subsection{Steady-state properties} 

We now turn to an analysis of the steady state beyond the classical limit, by comparing numerically exact results (obtained from the full master equation) to the semi-classical approximation of Eq.~(\ref{eq:LangevinEq}), which includes quantum fluctuations. Specifically, for the illustrative Cases I-IV, we display in Fig.~\ref{fig:Avr} the level-spacing ratio statistics $P(r)$ of the effective Hamiltonian, $H_\text{eff}$, obtained from the steady-state density matrix $\rho_0$. The statistics of the level-spacing ratios for the approximate classical density matrix $\rho_\text{C}$, obtained from Eq.~\eqref{eq:rho_Langevin_sample} at large times, are also given in Fig.~\ref{fig:Avr} as insets. Furthermore, we quantify in  Fig.~\ref{fig:EntropIPR}(a) the localization properties of the steady state by considering the finite-size scaling of the phase-space Inverse Participation Ratio, IPR$_\phi$, proposed in Eq.\eqref{eq:IPR_phi}. 
Finally, in Fig.~\ref{fig:EntropIPR}(b), we present the scaling of the steady-state entropy with $N$. 

For Case I, the level-spacing ratio statistics follows a Poissonian distribution, which is also well captured by the semi-classical density matrix, $\rho_\text{C}$. 
Both $\text{IPR}_\phi$ and $S$ are of order 1 and do not scale with $N$. 

Case II, for which the underlying classical dynamics is strongly chaotic, yields a $P(r)$ which si well described by the Gaussian Unitary Ensemble (GUE), for both $\rho_0$ and $\rho_\text{C}$. In this case,  $\text{IPR}_\phi \propto N^{-2} $ and $S \propto \ln N^{2} $. 

When the classical dynamics attains regular limiting cycles, as in Case III, we observe a Poissonian-like distribution $P(r)$, although  deviations are present for both $\rho_0$ and $\rho_\text{C}$. The magnitude of these deviations seems to decrease as $N$ increases. However, from the range of system sizes available to us, we could not conclude that $P(r)$ converges to a Poissonian distribution. Still, it is clear that no level repulsion is present, since $P(r)$ is maximal at the origin.   
Furthermore, although the finite-size scalings of $\text{IPR}_\phi$ and $S$ are not as conclusive as in previous cases, they are quite compatible with $\text{IPR}_\phi \propto N^{-1} $ and $S \propto \ln N^{1}$. 

In Case IV, when the classical dynamics shows mixed regular/chaotic features, also $P(r)$ has a mixed form, which can be reasonably fitted by a convex combination of Poissonian and GUE distributions. 
Although the numerical values of $\text{IPR}_\phi$ and $S$ are between Case II (fully chaotic) and and Case III (regular), the finite-size scaling for Case IV is not conclusive. Close to the upper limit of our simulations ($N=30$), both quantities seem to saturate with $N$, which likely indicates that the asymptotic scaling is not yet attained for these values of $N$.    

\subsection{Spectral Liouvillian properties } 

To better contextualize the discussion of the next section, we  discuss here some notable properties of the Liouvillian spectrum. Firstly, for all four representative Cases~I--IV, the distribution of complex level spacing ratios (CSR, recently introduced as a signature of dissipative chaos ~\cite{PhysRevX.Pedro}) diagnoses that the bulk of the spectrum is chaotic. We show these results in Appendix~\ref{App-F-Level-Statistics}.

Furthermore, we note that a non-stationary long-time trajectory implies the existence of multiple Liouvillian eigenvalues along the imaginary axis. 
However, for any finite $N$, we always find a unique steady state. These two facts can be easily reconciled if there is a number of eigenvalues approaching the imaginary axis as $N$ increases. 
In Appendix~\ref{App-E-Liouv-Gap} we corroborate this fact by presenting evidence that the Liouvillian gap $\Delta= - \text{Re} \Lambda_1 $ vanishes for cases II, III and IV in the large $N$ limit.

\section{Discussion} 
\label{sec:Discussion}

\begin{table*}
\begin{centering}
\begin{tabular}{|c|c|c|c|c|c|c|c|c|c|c|c|c|c|}
\hline 
\multicolumn{2}{|c|}{Classical dynamics} & \multicolumn{3}{c|}{Properties of $\mathcal{L}$} & \multicolumn{3}{c|}{Properties of $\rho_{0}$} & \multirow{2}{*}{Case} & \multicolumn{5}{c|}{Parameters}\tabularnewline
\cline{1-8}\cline{10-14}
Label & $\lambda\in\left[\lambda_{\text{min}},\lambda_{\text{max}}\right]$ & Symmetry & $\Delta$ & CSR & SR & $\text{IPR}_{\phi}$ & $S$ &  & $U$ & $w_{+1}$ & $w_{-1}$ & $w_{-2}$ & $w_{-3}$\tabularnewline
\hline 
\hline 
Attractor & $\lambda_{\text{max}}\lesssim 0$ & - & $N^{0}$ & Gin & P & $N^{0}$ & $\ln N^{0}$ & I & 1. & 1.4 &  & 1.1 & 1.2\tabularnewline
\hline 
Chaotic & $\lambda_{\text{min}}\gtrsim0$ & - & $N^{-1}$ & Gin & GUE & $N^{-2}$ & $\ln N^{2}$ & II & 6. &  & 1.1 & 1.2 & 1.4\tabularnewline
\hline 
Regular & $\lambda_{\text{min}}\simeq0$ & - & $N^{-1}$ & Gin & P & $N^{-1}$ & $\ln N^{1}$ & III & 0.01 &  & 0.1 & 0.5 & 0.4\tabularnewline
\hline 
Mixed & $\lambda_{\text{min}} \lesssim 0;\lambda_{\text{max}}\gtrsim0$ & - & $N^{-1}$ & Gin & Mixed & (?) & (?) & IV & 4. & 0.2 & 0.5 & 0.7 & 0.7\tabularnewline
\hline 
Regular (WS) & $\lambda_{\text{min}}\simeq0$ & Weak & $N^{-1}$ & P & $\propto1$ & $N^{-2}$ & $\ln N^{2}$ & V &  &  & 1.0 & 1.0 & 1.0\tabularnewline
\hline 
 &  & - & $N^{-1}$ & P & P & $N^{-2}$ & $\ln N^{2}$ & V' &  &  & 1.01 & 1.0 & 1.0\tabularnewline
\hline 
Chaotic (WS) & $\lambda_{\text{min}}\gtrsim0$ & Weak & $N^{-1}$ & Gin & $\propto1$ & $N^{-2}$ & $\ln N^{2}$ & VI & 6. &  & 1.0 & 1.0 & 1.0\tabularnewline
\hline 
 &  &  - & $N^{-1}$ & Gin & GUE & $N^{-2}$ & $\ln N^{2}$ & VI' &  &  & 1.01 & 1.0 & 1.0\tabularnewline
\hline 
Integrable & $\lambda_{\text{min}}\simeq0$ & Strong & $N^{-1}$ & P & P & $N^{-1}$ & $\ln N^{1}$ & VII{*} & 0.5 &  &  & 0.4 & 1.0\tabularnewline
\hline 
\end{tabular}
\par\end{centering}
\caption{Summary of results, illustrated by Cases I-IV, in the main text, and
Cases V-VII, shown in the Appendices. `Label' indicates the main
feature of the classical dynamics. $\lambda_{\text{min}}$ and $\lambda_{\text{max}}$
are extrema of the support of the Lyapunov exponent distribution.
`Symmetry' refers to existence of a weak, strong, or no (-) symmetry
of the Lindblad operator. The scalings with $N$ are provided for
the Lindblad gap ($\Delta$), the Phase-space Inverse Participation ratio
($\text{IPR}_{\phi}$), and the Entropy ($S$). Inconclusive finite-size
scaling is marked as (?). The Complex Spacing Ratio (CSR) signals if
the bulk spectrum of the Liouvilian is Possonian (P) or conforms to
the Ginibre ensemble distribution (Gin). The spacing ratios (SR),
marked as Poissonian (P) or conforming to that of Gaussian Unitary
Ensemble (GUE), refer to the statistics of the effective Hamiltonian
of the steady-state density matrix, $\rho_{0}.$ For all cases, $w_{+2}=w_{+3}=0$.
For cases $\text{I}-\text{VI}$, $J_{i=1,2,3}=1$. In case$\text{VII}^{*}$ we also set $J_{1(2)}=0$ and $J_{3}=1$, 
leading to a decoupled well. 
Empty entries correspond to zeroes.}
\label{tab:summary}
\end{table*}

The results from the previous section point to a clear quantum-classical correspondence in open quantum systems, using the $SU(3)$ Bose-Hubbard trimer as a representative example. A comprehensive summary of our findings is presented in Table~\ref{tab:summary}, which classifies the regimes analyzed according to classical dynamics, spectral properties of the Liouvillian ($\mathcal{L}$), and structural properties of the steady-state density matrix ($\rho_0$). 
This classification highlights clear distinctions between attractor, regular, chaotic, and mixed dynamical regimes.
Additional examples (Cases V–VII), discussed in detail in the Appendices, further support and complement the results of the previous section. Importantly, since no involutive symmetries -- such as time-reversal -- are imposed, all chaotic cases exhibit level statistics consistent with the Gaussian Unitary Ensemble.

\subsection{Steady-State Level Statistics }

A key finding of our analysis is the direct connection between the classical Lyapunov exponent and the level-spacing statistics of the effective Hamiltonian associated with the steady-state density matrix, $\rho_0$. Specifically, we observe two distinct scenarios: \begin{itemize} \item Systems exhibiting classical regular dynamics, characterized by non-positive Lyapunov exponents, yields steady states with Poissonian level statistics. \item Systems with classical chaotic dynamics, characterized by positive Lyapunov exponents, result in steady states whose spectral statistics conforms to the Gaussian Unitary Ensemble, indicative of quantum chaoticity. \end{itemize}

This correspondence between classical dynamics and quantum spectral statistics is robust across a wide range of regimes, though notable exceptions arise. 
One such case occurs when the steady state is proportional to the identity matrix -- a situation realized non-trivially in Cases~V and VI or, more commonly, when all jump operators are Hermitian. 
For the non-trivial Cases~V and VI, the condition $w_{+i} = w_+$ and $w_{-i} = w_-$ implies that the Lindblad operator possesses a weak symmetry~\cite{Buca_2012}, becoming invariant under well permutations.
In classical terms, this corresponds to the vanishing of the deterministic dissipative term in Eq.~(\ref{eq:X}), due to the condition $\mathcal{W}_{\mu i} = \bar{\mathcal{W}}_{\mu i}$ for Hermitian jump operators (or in the presence of the weak symmetry mentioned above). This cancellation, however, does not affect the noise term [see Eq.~(\ref{eq:M_matrix})], resulting in purely noise-driven dynamics. In analogy with fluctuation-dissipation relations in equilibrium systems, this corresponds to an infinite effective temperature, leading to a steady state with uniform phase-space distribution.
Importantly, any generic perturbation reintroduces a finite dissipative component, which distorts this uniform distribution. In the quantum description, this translates into lifting the degeneracy of the zero eigenvalues of $\rho_0$, causing the level statistics to align once more with the nature of the underlying classical dynamics -- Poissonian in regular regimes and GUE in chaotic ones, see Cases~V$'$ and VI$'$.

Another notable case is the regime with an attractive fixed point, where the spectrum of the effective Hamiltonian can be approximated analytically (see Appendix~\ref{app:attract}). Here, at leading order in $N$, the spectrum is a direct sum of two equally spaced grids,
$e_{\bs n}   = e_0 + c_2  n_{2} + c_3  n_{3} $,
where $n_{2,3}=0,1,2\ldots$ and $e_0$, $c_2$, $c_3$ are real constants.
Naturally, the direct sum of these two grids yields no intrinsic level repulsion. 
Introducing $1/N$ corrections breaks the equal-spacing property but does not introduce any qualitative correlations to the eigenvalues. This explains why the resulting statistics follows a Poisson distribution. 

Interestingly, while Lindblad operators exhibiting Poissonian CSR statistics are consistently associated with classical regular dynamics and steady-state density matrices displaying Poissonian (Hermitian) level statistics (e.g., Cases~V and VII$^*$), the converse does not necessarily hold. That is, Lindblad operators with fully chaotic spectra -- as indicated by Ginibre-type CSR -- can still exhibit regular long-time dynamics. This occurs trivially in situations dominated by a single attractive fixed point, such as Case~I, but also emerges in cases where the classical dynamics settles asymptotically into limit cycles, as seen in Case~III.

\subsection{Phase-space spreading in the steady state} 

To characterize how the quantum steady state occupies the classical phase space, we have introduced a phase-space inverse participation ratio (IPR$_\phi \propto N^{-D/2}$), defining an effective dimension $D$ for the support of the Husimi distribution. Our analysis reveals the following scaling behaviors:
\begin{itemize}
    \item For systems whose classical dynamics flows to an attractive fixed point, IPR$_\phi$ becomes independent of $N$, indicating a localized distribution ($D = 0$).
    \item When the asymptotic classical trajectories are non-stationary -- whether regular or chaotic -- and no strong symmetries are present, IPR$_\phi$ scales as $N^{-D/2}$ with $D$ equal to the full phase-space dimension (here, $D = 4$), reflecting maximal delocalization. This scaling is not modified by the presence of weak symmetries.
    \item In contrast, strong symmetries constrain the dynamics to lower-dimensional manifolds. When these symmetries are respected by the initial state, the effective phase-space dimension $D$ is reduced accordingly.
\end{itemize}

The last point is illustrated by Case~VII$^{*}$, where the particle number in well 1, $b^{\dagger}_1 b_1$, commutes with both the Hamiltonian and the jump operators. Assuming an initial state with a well-defined occupation of this mode, i.e., $b^{\dagger}_1 b_1 \rho_0 = n_1 \rho_0$, the dynamics remains confined to an $SU(2)$ submanifold of $SU(3)$. This removes one degree of freedom, yielding an effective dimension $D = 4 - 2 \times 1 = 2$, consistent with the scaling observed in Table~\ref{tab:summary}. Note that for this result to hold the initial state has to commute with the strong symmetry, otherwise the dynamics would spread across multiple symmetry sectors and the dimensional reduction would not take place.

It is also important to note that chaotic dynamics is ruled out when the effective phase-space dimension is reduced to $D = 2$. Therefore, in the trimer system ($D = 4$), any strong symmetry necessarily leads to regular Liouvillian dynamics and a regular steady state. However, in systems with higher dimensionality, such as four-well configurations, strong symmetries may still permit chaotic dynamics, albeit within reduced-dimensional manifolds.

Interestingly, Case~III exhibits an unexpected confinement to a lower-dimensional manifold ($D = 2$), despite the absence of identifiable strong symmetries and the presence of chaotic Liouvillian spectra. This behavior could indicate the existence of an approximate or hidden symmetry, although we were unable to identify any exact strong symmetry, even in the $U = 0$ limit.
An alternative explanation is that the observed scaling reflects finite-size effects. At larger values of $U$, the dynamics clearly flows to an attractor ($D = 0$) but, for small $N$, IPR$_\phi$ initially displays $N^{-1}$ scaling. It is only beyond a certain system size that IPR$_\phi$ saturates, becoming independent of $N$, consistent with localization at a fixed point. Within the accessible system sizes, we cannot rule out the possibility that such crossover behavior persists down to $U = 0$.
A final possibility is that the reduction in effective dimensionality arises from a genuinely dynamical mechanism that limits phase-space exploration -- independent of symmetry considerations -- which merits further investigation.

When the classical dynamics exhibits multiple coexisting behaviors -- for example, one region of phase space flows to an attractor while another exhibits chaos -- the effective dimension $D$ is expected to be determined by the component with the smallest support. This can be understood by considering the IPR$_\phi$ as a weighted sum over distinct dynamical components. Let $p_{(i)}$ denote the probability of the system occupying component $(i)$, and $\text{IPR}_\phi^{(i)}$ the corresponding inverse participation ratio. Then the total IPR is given by
$\text{IPR}_\phi = \sum_i p_{(i)}^2 \, \text{IPR}_\phi^{(i)}$, 
which is dominated by the term with the slowest decay, i.e., the smallest effective dimension.
In mixed dynamical regimes, such as Case~IV, this competition complicates the extraction of clean scaling exponents due to significant finite-size effects. Nevertheless, the curvature observed at large $N$ in Fig.~\ref{fig:EntropIPR} for Case~IV may signal the presence of an attracting sector within the phase space, which ultimately sets the asymptotic scaling of IPR$_\phi$.

\subsection{Classical features of the steady state}

Our study demonstrates the effectiveness of semiclassical methods -- specifically, stochastic classical mixtures of coherent states -- in capturing quantum steady-state properties beyond mean-field approximations. Remarkably, this approach accurately reproduces not only observable averages up to next-to-leading order in $1/N$, but also more subtle features such as entropy, level statistics, and phase-space delocalization.

The success of the semiclassical approximation underscores the fundamentally classical character of the steady states in this class of dissipative quantum systems. This classicality is most transparently revealed through the complementary diagnostics provided by the entropy $S$ and the phase-space inverse participation ratio, $\text{IPR}_\phi$. A pure quantum state ($S = 0$) that is strongly delocalized in phase space can only attain a small $\text{IPR}_\phi$ by forming a coherent superposition of many basis states, in contrast to classical mixtures where delocalization arises from statistical averaging.
For example, in the $SU(3)$ Bose-Hubbard trimer, pure states with macroscopically occupied modes -- such as $\ket{n_1 = N/3, n_2 = N/3, n_3 = N/3}$ -- exhibit $\text{IPR}_\phi \propto N^{-1}$, while random pure states display even stronger delocalization, with $\text{IPR}_\phi \propto N^{-2}$. By contrast, classical mixtures like the fully incoherent state $\rho_0 \propto \mathds{1}$ achieve low $\text{IPR}_\phi$ values by distributing classical probability broadly over phase space, without requiring quantum coherence.

Across all regimes explored in Table~\ref{tab:summary}, we find that the entropy consistently scales as $S \propto \ln N^D$, where $D$ is the effective dimension of the phase-space support. This scaling behavior mirrors that of classical probability distributions and provides compelling evidence for the accuracy of the semiclassical framework in capturing the structure and statistics of steady states in driven-dissipative quantum many-body systems.

Perhaps the most striking feature of the semiclassical approach is its ability to capture the level statistics of the quantum steady state. This can be understood by interpreting the spectral properties of $\rho_0$ through the semiclassical stochastic density matrix defined in Eq.~(\ref{eq:rho_Langevin_sample}).
When the classical dynamics is regular, coherent states evolve along well-structured, dissipative trajectories whose diffusion leads to a smooth and structured distribution in phase space. The resulting semiclassical steady state $\rho_\mathrm{C}$ exhibits limited eigenvalue correlations, consistent with Poissonian level statistics.
Conversely, chaotic classical dynamics generate highly sensitive and irregular trajectories, producing a fragmented and rough phase-space distribution. In the extreme limit, $\rho_\mathrm{C}$ resembles an incoherent average over uncorrelated random vectors, mimicking the statistics of random matrix ensembles. This correspondence provides an intuitive explanation for the emergence of Wigner-Dyson (GUE) level statistics in chaotic regimes and Poissonian statistics in regular ones.

\section{Conclusion }
\label{sec:Conclusion}

In this work, we explored the quantum–classical correspondence in open quantum systems through the $SU(3)$ Bose-Hubbard trimer -- a minimal yet remarkably rich model. By combining exact numerical methods with a semiclassical stochastic framework, we uncovered how distinct classical dynamical regimes -- fixed-point attractors, limit cycles, and chaotic trajectories -- are faithfully reflected in the spectral and structural properties of the quantum steady state.

Our key finding is that classical dynamics (notably, the sign of the Lyapunov exponent) determines the spectral statistics of the steady state's effective Hamiltonian, $H_{\text{eff}}$: non-positive exponents (regular dynamics) lead to Poissonian level statistics and localized, while positive exponents (chaotic dynamics) yield Wigner-Dyson statistics, delocalized Husimi distributions, and high entropy -- hallmarks of quantum ergodicity.

To quantify localization, we introduced the phase-space inverse participation ratio IPR$_\phi$, which reveals an effective dimension $D$ of the steady state’s classical support. Attractors yield $D=0$ (localized states), while non-stationary dynamics without strong symmetries yield $D=4$ (maximal delocalization). Strong symmetries constrain the dynamics and reduce $D$, precluding chaos in low-dimensional systems. Intriguingly, we identified regimes with apparent dynamical confinement to lower-dimensional manifolds without clear symmetry protection, hinting at either hidden structure or strong finite-size effects.

The semiclassical description -- based on coherent-state mixtures -- proved remarkably accurate, reproducing entropy scaling, spectral statistics, and phase-space features beyond leading-order observables. This success underscores the classical nature of the steady state in large-$N$ dissipative quantum systems, with the entropy consistently scaling as $S \propto \ln N^D$ across all regimes.

Our findings provide a semiclassical explanation for the integrability–chaos dichotomy in steady-state level statistics, as originally observed by Prosen and \v{Z}nidari\v{c}~\citep{prosen2013eigenvalue} in one-dimensional chains with no classical counterpart. Moreover, we clarify an important subtlety in the broader spectral characterization of open quantum systems: while Liouvillians exhibiting regular spectral statistics -- as diagnosed by the complex spacing ratio~\citep{sa2020complex} -- consistently yield steady states with Poissonian $H_{\text{eff}}$ statistics, Liouvillians conforming to Ginibre (GinUE) spectral statistics can give rise to steady states that are regular, chaotic, or mixed in terms of their internal eigenvalue structure. This highlights the crucial distinction between the spectral properties of the full Liouvillian and those of the effective Hamiltonian encoding the steady-state structure, and helps clarify which signatures of dissipative quantum chaos are ultimately encoded in the steady state itself.

The richness of this minimal model allows us to address a central open question in the theory of dissipative quantum systems with a classical limit: Are semiclassical and quantum chaos encoded in the steady-state density matrix? Our answer is affirmative. Classical chaos translates into quantum chaotic signatures in steady-state spectra and phase-space diagnostics, offering a systematic framework for interpreting quantum ergodicity in open systems.

Nonetheless, we caution that the semiclassical approximation may break down in non-generic cases, such as critical steady states associated with diverging fluctuations or non-equilibrium phase transitions~\cite{RABlythe_2006}. While such regimes are typically destabilized by small perturbations, they remain important targets for future work. In particular, the tools developed here -- such as IPR$_\phi$ and the classical density matrix -- provide a natural lens to investigate the correspondence between the quantum and classical perspectives.

In summary, our study demonstrates that signatures of classical dynamics -- integrable or chaotic -- are robustly encoded in the spectral and structural features of steady states in dissipative quantum systems. These findings establish a solid foundation for using semiclassical tools to analyze quantum chaos and ergodicity in open, interacting systems of experimental relevance.

\emph{Added Note.---}
We note a recent preprint by Mondal, Santos, and Sinha~\cite{mondal2025transientsteadystatechaosdissipative} that examines transient and steady‑state chaos in the open anisotropic Dicke model. Using entanglement growth and out-of-time-order correlators (OTOCs), they identify both early‑time and long‑time chaotic dynamics. Their findings are in good agreement with ours, when the diagnostic framework overlaps, and are furthermore complementary, given the different model systems considered.

\acknowledgements

This work was supported by FCT-Portugal, Grant Agreement No. 101017733 (PR), as part of the QuantERA II project “DQUANT: A Dissipative Quantum Chaos perspective on Near-Term Quantum Computing”~\footnote{\url{https://doi.org/10.54499/QuantERA/0003/2021}}. 
PR acknowledges further support from FCT through the financing of the I\&D unit: UID/04540 - Centro de Física e Engenharia de Materiais Avançados. 
SC acknowledges support from the Innovation Program for Quantum Science and Technology (Grant No.~2021ZD0301602) and the National Science Association Funds (Grant No.~U2230402).

\clearpage 

\appendix
\section{$\boldsymbol{SU(3)}$ generators \label{App-A-SU3-Gen}}

Defining the generators of $SU(3)$ as 
\begin{align}\label{HEdef}
H_{1}= & b_{3}^{\dagger}b_{3}-b_{2}^{\dagger}b_{2},\nonumber \\
H_{2}= & b_{2}^{\dagger}b_{2}-b_{1}^{\dagger}b_{1},\nonumber \\
E_{1}= & b_{3}^{\dagger}b_{2},\nonumber \\
E_{2}= & b_{2}^{\dagger}b_{1},\nonumber \\
E_{3}= & b_{3}^{\dagger}b_{1},
\end{align}
it is straightforward to check that they obeying the $SU(3)$ commutation
relations in the Chevalley basis~\cite{Guaita2021,Gilmore1974,Perelomov1986}:
\begin{align*}
[H_{i},E_{\pm j}]= & \pm\alpha_{i}^{(j)}E_{\pm j},\\
\left[E_{j},E_{-j}\right]= & H_{j}\\
\left[E_{\pm i},E_{\pm j}\right]= & \pm E_{\pm k},
\end{align*}
for $\left(i,j,k\right)=\left(1,2,3\right)$ and cyclic permutations
of the indices, and where $\boldsymbol{\alpha}^{(1)}=(2,-1)^{T}$,
$\boldsymbol{\alpha}^{(2)}=(-1,2)^{T}$ and $\boldsymbol{\alpha}^{(3)}=(1,1)^{T}$,
while $E_{-j}=E_{j}^{\dagger}$. 

In this basis, the Hamiltonian in Eq.~(\ref{Eq:Hamiltonian}) becomes
\begin{align}
H= & -J\left(E_{1}+E_{2}+E_{3}+\text{h.c.}\right)\\
+ & \frac{U}{3N}\left[H_{1}^{2}+H_{1}H_{2}+H_{2}^{2}+\frac{1}{2}N\left(N-3\right)\right].\nonumber 
\end{align}
and the jump operators are given by 
\begin{equation}
W_{-1}=\sqrt{w_{-1}}E_2,\ W_{-2}=\sqrt{w_{-2}}E_1,\  W_{-3}=\sqrt{w_{-3}}E_3^{\dagger}
\end{equation}
with $W_{+i}=W_{-i}^{\dagger}$.

\section{$\boldsymbol{SU(3)}$ coherent states \label{App-B-SU3-CS}}

The coherent states, defined as follows, are parameterized by the
two complex numbers $\eta_{2}$ and $\eta_{3}$: 
\begin{equation}
\ket{\boldsymbol{\eta}}=e^{\boldsymbol{E}\cdot\boldsymbol{\eta}}\ket{N,0,0},
\end{equation}
where $\boldsymbol{E}\cdot\boldsymbol{\eta}=\eta_{2}E_{2}+\eta_{3}E_{3}$
and $\ket{N,0,0}$ is the reference state with $N$ bosons in well
1. Using the definition of $E_{2}$ and $E_{3}$, given in Eq.~(\ref{HEdef}),
the inner product between two coherent states is obtained as 
\begin{eqnarray}
\braket{\boldsymbol{\eta}}{\boldsymbol{\eta}^{\prime}} & = & (1+\bar{\eta}_{2}\eta_{2}^{\prime}+\bar{\eta}_{3}\eta_{3}^{\prime})^{N}=(1+\boldsymbol{\eta}^{\dagger}\cdot\boldsymbol{\eta}^{\prime})^{N},
\end{eqnarray}
where here and in the following we indicate complex conjugation with
a bar, i.e., $\bar{\eta}_{i}=\eta_{i}^{*}$. The coherent states form
an over-complete basis~\cite{TViscondia2011} satisfying: 
\begin{equation}
\int d\mu\left(\boldsymbol{\eta}^{\dagger},\boldsymbol{\eta}\right)\frac{\ket{\boldsymbol{\eta}}\bra{\boldsymbol{\eta}}}{\braket{\boldsymbol{\eta}}{\boldsymbol{\eta}}}=1,\label{Eq-21}
\end{equation}
where for $SU(3)$ the measure is defined as $d\mu\left(\boldsymbol{\eta}^{\dagger},\boldsymbol{\eta}\right)=\frac{(N+1)(N+2)}{\pi^{2}(1+\boldsymbol{\eta}^{\dagger}\cdot\boldsymbol{\eta})^{3}}d(\text{Re}\eta_{2})d(\text{Im}\eta_{2})d(\text{Re}\eta_{3})d(\text{Im}\eta_{3})$.

To facilitate describing the evolution of coherent states, it is convenient
to rewrite the generator of the $SU(3)$ algebra in a differential form.
By using a Baker-Campbell-Hausdorff-like formula for $SU(3)$~\cite{Weigert},
the following differential representation is obtained: 
\begin{eqnarray}
H_{1} & = & -\eta_{2}\partial_{\eta_{2}}+\eta_{3}\partial_{\eta_{3}}\\
H_{2} & = & -n+2\eta_{2}\partial_{\eta_{2}}+\eta_{3}\partial_{\eta_{3}}\\
E_{1} & = & \eta_{2}\partial_{\eta_{3}}\\
E_{1}^{\dagger} & = & \eta_{3}\partial_{\eta_{2}}\\
E_{2} & = & \partial_{\eta_{2}}\\
E_{2}^{\dagger} & = & n\eta_{2}-\eta_{2}^{2}\partial_{\eta_{2}}-\eta_{2}\eta_{3}\partial_{\eta_{3}}\\
E_{3} & = & \partial_{\eta_{3}}\\
E_{3}^{\dagger} & = & n\eta_{3}-\eta_{3}^{2}\partial_{\eta_{3}}-\eta_{2}\eta_{3}\partial_{\eta_{2}}.
\end{eqnarray}
These differential expressions are useful to compute expectation values
of the generators, for example; 
\begin{equation}
\frac{\bra{\boldsymbol{\eta}}E_{1}\ket{\boldsymbol{\eta}}}{\braket{\boldsymbol{\eta}}{\boldsymbol{\eta}}}=\frac{1}{\braket{\boldsymbol{\eta}}{\boldsymbol{\eta}}}\eta_{2}\partial_{\eta_{3}}\braket{\boldsymbol{\eta}}{\boldsymbol{\eta}}=\frac{N\eta_{2}\bar{\eta}_{3}}{1+|\eta_{2}|^{2}+|\eta_{3}|^{2}}.
\end{equation}
Note that, as usual, $\eta_{i}$ and $\bar{\eta}_{i}$ should be treated
as independent variables (i.e. $\partial\bar{\eta}_{i}/\partial\eta_{i}=0$).

\section{Semi-classical Dissipative Dynamics \label{App-C-SC-DissD}}

At time $t=T$, the time-evolved density matrix, $\rho\left(T\right)$,
can be obtained from its coherent state matrix elements 
\begin{equation}
\mathcal{Z}_{T}=\bra{\boldsymbol{\eta}'}\rho\left(t\right)\ket{\boldsymbol{\eta}}=\bra{\bra{\boldsymbol{\eta},\bar{\boldsymbol{\eta}}'}}\mathrm{T} e^{\int_{0}^{t}\mathcal{L}_{t^{\prime}}dt^{\prime}}\ket{\ket{\rho(0)}}
\end{equation}
with $\mathrm{T} $ the time-ordering operator and $\ket{\ket{\boldsymbol{\eta},\bar{\boldsymbol{\eta}}'}}$
the double-contour coherent-state in the vectorized notation 
\begin{equation}
\ket{\ket{\boldsymbol{\eta},\bar{\boldsymbol{\eta}}'}}=\ket{\boldsymbol{\eta}}\otimes\bra{\boldsymbol{\eta}^{\prime}}^{T}=\ket{\boldsymbol{\eta}}\otimes\ket{\bar{\boldsymbol{\eta}}^{\prime}}.\label{Eq-doubled-ket}
\end{equation}
This propagator can be evaluated by Trotter slicing in time $\Delta t=T/M$
and inserting a set of overcomplete coherent-states between each time
slice. The vectorized version of the completeness relation follows
from Eq.~(\ref{Eq-21}) 
\begin{eqnarray}
\int d\mu\left(\bs{\eta}{}_{\tau}^{\dagger+},\bs{\eta}_{\tau}^{+}\right)d\mu\left(\bs{\eta}{}_{\tau}^{\dagger-},\bs{\eta}_{\tau}^{-}\right)\frac{\ket{\ket{\boldsymbol{\eta}_{\tau}^{+},\boldsymbol{\eta}_{\tau}^{-}}}\bra{\bra{\boldsymbol{\eta}_{\tau}^{+},\boldsymbol{\eta}_{\tau}^{-}}}}{\langle\langle\boldsymbol{\eta}_{\tau}^{+},\boldsymbol{\eta}_{\tau}^{-}||\boldsymbol{\eta}_{\tau}^{+},\boldsymbol{\eta}_{\tau}^{-}\rangle\rangle} & = & \mathds{1}, \nonumber \\
\end{eqnarray}
where $\boldsymbol{\eta}_{\tau}^{\pm}$ stand for the forward and
backward fields of the $\tau$-th slice of the Keldysh contour. 

After Trotter slicing with the partition function above, we obtain
\begin{eqnarray}
\mathcal{Z}_{t} & = &  \bra{\bra{\boldsymbol{\eta},\boldsymbol{\eta}}}\cdots\mathds{1}_{2}e^{\Delta t\mathcal{L}_{\tau}}\mathds{1}_{1}e^{\Delta t\mathcal{L}_{\tau}}\mathds{1}_{0}\ket{\ket{\rho(0)}}  \nonumber \\
  & \simeq \int D\eta & e^{ \Delta t \sum_\tau \mathcal{S}_{\tau}}, \label{Z1}
\end{eqnarray}
where $D\eta = \prod_{\tau=0}^{M-1} d\mu\left(\bs{\eta}{}_{\tau}^{\dagger+},\bs{\eta}_{\tau}^{+}\right)d\mu\left(\bs{\eta}{}_{\tau}^{\dagger-},\bs{\eta}_{\tau}^{-}\right)$
and 
\begin{eqnarray}
 & & e^{ \Delta t \, \mathcal{S}_{\tau}} = \\
 & & \frac{\bra{\bra{\boldsymbol{\eta}_{\tau+1}^{+},\boldsymbol{\eta}_{\tau+1}^{-}}}e^{\Delta t\mathcal{L}_{\tau}}\ket{\ket{\boldsymbol{\eta}_{\tau}^{+},\boldsymbol{\eta}_{\tau}^{-}}}}{\sqrt{\langle\langle\boldsymbol{\eta}_{\tau+1}^{+},\boldsymbol{\eta}_{\tau+1}^{-}||\boldsymbol{\eta}_{\tau+1}^{+},\boldsymbol{\eta}_{\tau+1}^{-}\rangle\rangle\langle\langle\boldsymbol{\eta}_{\tau}^{+},\boldsymbol{\eta}_{\tau}^{-}||\boldsymbol{\eta}_{\tau}^{+},\boldsymbol{\eta}_{\tau}^{-}}\rangle\rangle}.\nonumber 
 \label{exl}
\end{eqnarray}

Using a smooth-path hypothesis,  
\begin{equation}
\boldsymbol{\eta}_{\tau+1}^{\pm} \simeq \boldsymbol{\eta}^{\pm}(t)+\Delta t \,\partial_{t}\boldsymbol{\eta}^{\pm}(t),
\end{equation}
and taking the limit of Trotter slices $M\to\infty$, the generating function becomes 
\begin{equation}
\mathcal{Z}_{t}=\int D\eta\,e^{iN\mathcal{S}},\label{Z1}
\end{equation}
with Keldysh  action given by 
\begin{widetext}
\begin{eqnarray}\label{Action}
\mathcal{S} & \simeq & -\frac{i}{2}\int d\tau\left[-\partial_{\boldsymbol{\eta}^{+}}\Omega^{+}\partial_{\tau}\boldsymbol{\boldsymbol{\eta}}^{+}+\partial_{\bar{\boldsymbol{\eta}}^{+}}\Omega^{+}\partial_{\tau}\bar{\boldsymbol{\eta}}^{+}-\partial_{\bar{\boldsymbol{\eta}}^{-}}\Omega^{-}\partial_{\tau}\bar{\boldsymbol{\eta}}^{-}+\partial_{\boldsymbol{\eta}^{-}}\Omega^{-}\partial_{\tau}\boldsymbol{\eta}^{-}\right]\nonumber \\
 & + & \int d\tau\sum_{\ell}\left[-\left(\mathcal{H}^{+}-\frac{i}{2}\mathcal{\bar{W}}_{\ell}^{+}\mathcal{W}_{\ell}^{+}\right)+\left(\mathcal{H}^{-}+\frac{i}{2}\mathcal{\bar{W}}_{\ell}^{-}\mathcal{W}_{\ell}^{-}\right)-i\mathcal{W}_{\ell}^{+}\mathcal{\bar{W}}_{\ell}^{-}+\frac{i}{2N}\left(\Delta\mathcal{W}_{\ell}^{+}+\Delta\mathcal{W}_{\ell}^{-}\right)\right].
\end{eqnarray}
The first term is the dynamical Berry phase with $\Omega^\pm = N^{-1} \ln \braket{\boldsymbol{\eta}^\pm}{\boldsymbol{\eta}^\pm} $. 
The Hamiltonian matrix elements are defined as
\begin{equation}\label{W_Wbar_def}
 \mathcal{H}^\pm=\frac{1}{N}\frac{\bra{\boldsymbol{\eta}^\pm}H\ket{\boldsymbol{\eta}^\pm}}{\braket{\boldsymbol{\eta}^\pm}{\boldsymbol{\eta}^\pm}}.
\end{equation}
Finally, the dissipation is described by the matrix elements of the jump operators 
\begin{equation}
\mathcal{W}_{\ell}^\pm =\frac{1}{N}\frac{\bra{\boldsymbol{\eta}^\pm}W_{\ell}\ket{\boldsymbol{\eta}^\pm}}{\braket{\boldsymbol{\eta}^\pm}{\boldsymbol{\eta}^\pm}},  \ \ \   \bar{\mathcal{W}}_{\ell}^\pm =\frac{1}{N}\frac{\bra{\boldsymbol{\eta}^\pm}W_{\ell}^\dagger\ket{\boldsymbol{\eta}^\pm}}{\braket{\boldsymbol{\eta}^\pm}{\boldsymbol{\eta}^\pm}},
\end{equation}
and their subdominant $1/N$ corrections 
\begin{eqnarray}
\Delta\mathcal{W}_{\ell}^\pm & = & \frac{1}{N}\left[\frac{\bra{\boldsymbol{\eta}^\pm}W_{\ell}^{\dagger}W_{\ell}\ket{\boldsymbol{\eta}^\pm}}{\braket{\boldsymbol{\eta}^\pm}{\boldsymbol{\eta}^\pm}}  
-\frac{\bra{\boldsymbol{\eta}^\pm}W_{\ell}^{\dagger}\ket{\boldsymbol{\eta}^\pm}}{\braket{\boldsymbol{\eta}^\pm}{\boldsymbol{\eta}^\pm}}\frac{\bra{\boldsymbol{\eta}^\pm}W_{\ell}\ket{\boldsymbol{\eta}^\pm}}{\braket{\boldsymbol{\eta}^\pm}{\boldsymbol{\eta}^\pm}}\right],
\end{eqnarray}

For large-$N$ the saddle-point condition, $\delta_{\bar{\eta}^{\pm}}\mathcal{S}=\delta_{\eta^{\pm}}\mathcal{S}=0$, 
admits a solution where $\eta^{q} = \eta^{+}-\eta^{-} = 0$, and $\eta^c = (\eta^{+}+\eta^{-})/2$, where
\begin{eqnarray}
\delta_{\boldsymbol{\bar\eta}^q  }\mathcal S & = & 
 \boldsymbol{\omega} \partial_{\tau}\boldsymbol{\eta}^c
+i \partial_{\Bar{\boldsymbol{\eta}}}\mathcal{H}
+\frac{1}{2} \sum_{\ell} \mathcal{W}_{\ell} \partial_{\Bar{\boldsymbol{\eta}}}\mathcal{\bar{W}}_{\ell}
-\frac{1}{2} \sum_{\ell}\mathcal{\bar{W}}_{\ell}\partial_{\Bar{\boldsymbol{\eta}}}\mathcal{W}_{\ell},\label{EqMot1}\\
\delta_{\boldsymbol{\eta}^q  }\mathcal S
& = & 
- \boldsymbol{\omega}^{T}\partial_{\tau}\Bar{\boldsymbol{\eta}}^c 
+i\partial_{\boldsymbol{\eta}}\mathcal{H}-\frac{1}{2}\sum_{\ell}\mathcal{\bar{W}}_{\ell}\partial_{\boldsymbol{\eta}}\mathcal{W}_{\ell}+\frac{1}{2}\sum_{\ell}\mathcal{W}_{\ell}\partial_{\boldsymbol{\eta}}\mathcal{\bar{W}}_{\ell},\label{EqMot2}
\end{eqnarray}

where the matrix elements of $\boldsymbol{\omega}$ are given by $\omega_{j,k}=N^{-1}\partial_{\bar{\eta}_{j}}\partial_{\eta_{k}}\ln
\braket{\boldsymbol{\eta}}{\boldsymbol{\eta}}$. 
\end{widetext}

\section{Stochastic classical process}
\label{App-D-StocCP}
Here, we consider quantum fluctuations around the $\eta^q =0 $ saddle-point derived in Sec.~\ref{App-C-SC-DissD}. These fluctuations yield a $1/N$ correction through an effective stochastic Langevin term  \cite{kamenev2011fieldtheory}. 

To quadratic order in $\eta^q =0 $, after applying a Keldysh rotation, the action of Eq.~(\ref{Action}) becomes 
\begin{equation}
\label{EqSexpandedc}
\mathcal{S}\simeq \int d\tau
\left\{  {\begin{pmatrix}
\boldsymbol{\bar{\eta}}^{q} \\
\boldsymbol{\eta}^{q}
\end{pmatrix}}^{T}
\begin{pmatrix}
\delta_{\boldsymbol{\bar{\eta}}^{q}} \mathcal S 
\\ 
\delta_{\boldsymbol{\eta}^{q}} \mathcal S 
\end{pmatrix}
 +\frac{1}{2}{\begin{pmatrix}
\boldsymbol{\bar{\eta}}^{q} \\
\boldsymbol{\eta}^{q}
\end{pmatrix}}^{T}
M
\begin{pmatrix}
\boldsymbol{\eta}^{q} \\
\boldsymbol{\bar{\eta}}^{q}
\end{pmatrix} \right\}
\end{equation}
with
\begin{eqnarray} \label{eq:M_matrix}
M_{j,k}&=&\frac{1}{2}\sum_\ell \left(
\partial_{X_j}\mathcal{W}_{\ell}^{c}
\partial_{\bar X_k}\mathcal{\bar{W}}_{\ell}^{c}+
\partial_{ \bar X_k}\mathcal{W}_{\ell}^{c}
\partial_{X_j}\mathcal{\bar{W}}_{\ell}^{c}\right)  
\end{eqnarray}
with $X =(\eta_2,\eta_3,\bar\eta_2,\bar\eta_3)$.

We follow the standard procedure \cite{kamenev2011fieldtheory} and make use of the complex Gaussian identity~\cite{Altland} 
\begin{widetext}
$$(2\pi)^{d}\det(A)e^{-\frac{1}{2}\left[\begin{pmatrix}\boldsymbol{\bar{\eta}}^{q}\\
\boldsymbol{\eta}^{q}
\end{pmatrix}^{T}A\begin{pmatrix}\boldsymbol{\eta}^{q}\\
\boldsymbol{\bar{\eta}}^{q}
\end{pmatrix}\right]}	=\int Dz\,\,e^{\left[-\frac{1}{2}\begin{pmatrix}\boldsymbol{\bar{z}}\\
\boldsymbol{z}
\end{pmatrix}^{T}(A)^{-1}\begin{pmatrix}\boldsymbol{z}\\
\boldsymbol{\bar{z}}
\end{pmatrix}+i\begin{pmatrix}\boldsymbol{\bar{\eta}}^{q}\\
\boldsymbol{\eta}^{q}
\end{pmatrix}^{T}\cdot\begin{pmatrix}\boldsymbol{z}\\
\boldsymbol{\bar{z}}
\end{pmatrix}\right]}$$    
to decouple the quadratic quantum $\boldsymbol{\eta}$ term in Eq.~(\ref{EqSexpandedc}). This yields
\begin{eqnarray}
Z&=&\int D\eta\  e^{iN \mathcal{S} }
\simeq
\int D\eta e^{iN\int d\tau\left\{ \begin{pmatrix}\boldsymbol{\bar{\eta}}^{q}\\
\boldsymbol{\eta}^{q}
\end{pmatrix}^{T}\begin{pmatrix}\delta_{\boldsymbol{\bar{\eta}}^{q}}\mathcal{S}\\
\delta_{\boldsymbol{\eta}^{q}}\mathcal{S}
\end{pmatrix}\right\} }\int Dz\,\,e^{\left[-\frac{1}{2}\begin{pmatrix}\boldsymbol{\bar{z}}\\
\boldsymbol{z}
\end{pmatrix}^{T}M^{-1}\begin{pmatrix}\boldsymbol{z}\\
\boldsymbol{\bar{z}}
\end{pmatrix}+i\sqrt{N}\begin{pmatrix}\boldsymbol{\bar{\eta}}^{q}\\
\boldsymbol{\eta}^{q}
\end{pmatrix}^{T}\begin{pmatrix}\boldsymbol{z}\\
\boldsymbol{\bar{z}}
\end{pmatrix}\right]}\\ \nonumber
&=&\int Dz\,\,e^{-\frac{1}{2}\begin{pmatrix}\boldsymbol{\bar{z}}\\
\boldsymbol{z}
\end{pmatrix}^{T}M^{-1}\begin{pmatrix}\boldsymbol{z}\\
\boldsymbol{\bar{z}}
\end{pmatrix}}\delta\left\{ \begin{pmatrix}\delta_{\boldsymbol{\bar{\eta}}^{q}}\mathcal{S}\\
\delta_{\boldsymbol{\eta}^{q}}\mathcal{S}
\end{pmatrix}+\frac{1}{\sqrt{N}}\begin{pmatrix}\boldsymbol{z}\\
\boldsymbol{\bar{z}}
\end{pmatrix}\right\}
\end{eqnarray}
where the last equality was obtained by integrating $\boldsymbol{\eta}^{q}$. 
The resulting porcess corresponds to a Langevin dynamics obtained from adding a stochastic noise term to the classical equations of motion. 

To preserve the symplectic form of the stochastic dynamics, the noise correlation matrix, $M$, can be diagonalized by a symplectic transformation $R$, that keeps the symplectic form $J=\text{diag}\left(1,1,-1,-1\right)$ invariant, i.e. $RJR^{\dagger}=J$. Therefore we can write 
\begin{equation}
M=R^{-1}\left(JRJMR^{\dagger}\right)R^{\dagger-1}=R^{-1}\left(JD\right)R^{\dagger-1}  , 
\end{equation}
where $JD=\text{diag}\left(\epsilon_{1},\epsilon_{2},\epsilon_{1},\epsilon_{2}\right)$, with $\epsilon_{i}>0$, is a positive-defined matrix. Changing the integration variables to
\begin{equation}
 \begin{pmatrix}\boldsymbol{z}\\
\boldsymbol{\bar{z}}
\end{pmatrix}	=R^{-1}\left(JD\right)^{1/2}\begin{pmatrix}\boldsymbol{\xi}\\
\boldsymbol{\bar{\xi}}
\end{pmatrix} , \label{stocahsticnoise}  
\end{equation}
we obtain 
\begin{equation}
    Z	=\ \int D\xi\,\,e^{-\bs{\xi}^{\dagger}\bs{\xi}}\delta\left[\begin{pmatrix}\delta_{\boldsymbol{\bar{\eta}}^{q}}\mathcal{S}\\
\delta_{\boldsymbol{\eta}^{q}}\mathcal{S}
\end{pmatrix}+\frac{1}{\sqrt{N}}\boldsymbol{G}\begin{pmatrix}\boldsymbol{\xi}\\
\boldsymbol{\bar{\xi}}
\end{pmatrix}\right],
\end{equation}
where $\boldsymbol{G}=R^{-1}\left(JD\right)^{1/2}$. 
The Langevin equations thus reads 
\begin{eqnarray}
  \partial_{\tau}\boldsymbol{\eta}^{c}&=&	-i\omega^{-1}\partial_{\bar{\boldsymbol{\eta}}}\mathcal{H}-\frac{1}{2}\sum_{\ell}\omega^{-1}\left(\bar{\mathcal{W}}_{\ell}\partial_{\bar{\boldsymbol{\eta}}}\mathcal{W}_{\ell}-\mathcal{W}_{\ell}\partial_{\bar{\boldsymbol{\eta}}}\bar{\mathcal{W}}_{\ell}\right)+i\frac{1}{\sqrt{N}}\omega^{-1}\boldsymbol{z}\\ 
\partial_{\tau}\bar{\boldsymbol{\eta}}^{c}&=&i(\omega^{-1})^  {T}\partial_{\boldsymbol{\eta}}\mathcal{H}-\frac{1}{2}\sum_{\ell}(\omega^{-1})^  {T}\left(\bar{\mathcal{W}}_{\ell}\partial_{\boldsymbol{\eta}}\mathcal{W}_{\ell}-\mathcal{W}_{\ell}\partial_{\boldsymbol{\eta}}\bar{\mathcal{W}}_{\ell}\right)-i\frac{1}{\sqrt{N}}(\omega^{-1})^  {T}\bar{\boldsymbol{z} }, 
\end{eqnarray}
where the stochastic noise $\boldsymbol{z}$ is defined by Eq.~(\ref{stocahsticnoise}) with $\xi$ a normal distributed complex stochastic variable with expectation values $  \mathbb{E}[ \bar{\xi}_{i}\left(t\right)\xi_{j}\left(t'\right) ] =\delta_{ij}\delta\left(t-t'\right)$.

\section{Spectrum of the Steady State for the Case of an Attractive Fixed Point} \label{app:attract}

Near an attractive fixed point the equations of motion can be linearized.
For simplicity, and without lost of generality, we assume the fixed
point is at $\boldsymbol{\eta}=0$, for which case 
\begin{align*}
\pd_{\tau}\begin{pmatrix}\boldsymbol{\eta}\\
\boldsymbol{\bar{\eta}}
\end{pmatrix}=- & \bs{\Lambda}\begin{pmatrix}\boldsymbol{\eta}\\
\boldsymbol{\bar{\eta}}
\end{pmatrix}-\frac{1}{\sqrt{N}}\begin{pmatrix}\boldsymbol{z}\\
\boldsymbol{\bar{z}}
\end{pmatrix},
\end{align*}
with 
\begin{align*}
\mathbb{E}\left[\begin{pmatrix}\boldsymbol{z}\left(t\right)\\
\boldsymbol{\bar{z}}\left(t\right)
\end{pmatrix}\begin{pmatrix}\boldsymbol{z}\left(t'\right)\\
\boldsymbol{\bar{z}}\left(t'\right)
\end{pmatrix}^{\dagger}\right] & =\bs M\delta\left(t-t'\right),
\end{align*}
and where both $\bs{\Lambda}$ and $\bs M$ are constant matrices.
By stability of the fixed point, the eigenvalues of $\bs{\Lambda}$
have a positive real part, and $\bs M$ is a positive defined hermitian
matrix that characterizes the noise. Solving explicitly the equations
of motion, we have that for large times 
\begin{align*}
\mathbb{E}\left[\begin{pmatrix}\boldsymbol{\eta}\left(t\right)\\
\boldsymbol{\bar{\eta}}\left(t\right)
\end{pmatrix}\begin{pmatrix}\boldsymbol{\eta}\left(t\right)\\
\boldsymbol{\bar{\eta}}\left(t\right)
\end{pmatrix}^{\dagger}\right] & =\frac{1}{N}\tilde{\bs M},
\end{align*}
where 
\begin{align*}
\tilde{\bs M} & =\sum_{\alpha\alpha'}\frac{\ket{\alpha}\bra{\tilde{\alpha}}\bs M\ket{\tilde{\alpha}'}\bra{\alpha'}}{\Lambda_{\alpha}+\bar{\Lambda}_{\alpha'}}
\end{align*}
is a positive defined hermitian matrix, with $\ket{\alpha}$ and $\bra{\tilde{\alpha}}$
the right and left eigenvalues of $\bs{\Lambda}$ with eigenvalue
$\Lambda_{\alpha}$. Working in the basis where $\bs{\tilde{M}}$
is diagonal, i.e. $\bs{\tilde{M}}=\text{diag\ensuremath{\left\{  \varepsilon_{2},\varepsilon_{3},\varepsilon_{2},\varepsilon_{3}\right\} } }$,
we get 
\begin{align*}
\rho_{0} & \propto\int d\eta_{2}d\eta_{3}\,e^{-N\left(\varepsilon_{2}\bar{\eta}_{2}\eta_{2}+\varepsilon_{3}\bar{\eta}_{3}\eta_{3}\right)}\frac{\ket{\boldsymbol{\eta}}\bra{\boldsymbol{\eta}}}{\braket{\boldsymbol{\eta}}{\boldsymbol{\eta}}}.
\end{align*}
Using the overlap of the coherent states with the number basis 
\begin{align*}
\braket{\bs n}{\boldsymbol{\eta}} & =\sqrt{\frac{N!}{n_{1}!n_{2}!n_{3}!}}\frac{\eta_{2}^{n_{2}}\eta_{3}^{n_{3}}}{\left(1+\abs{\eta_{2}}^{2}+\abs{\eta_{3}}^{2}\right)^{N/2}},
\end{align*}
one obtains
\begin{align*}
\bra{\bs n}\rho_{0}\ket{\bs n'} & \propto\delta_{ \bs n, \bs n'}\int d\eta_{2}d\eta_{3}\,e^{-N\left(\varepsilon^{-1}_{2}\bar{\eta}_{2}\eta_{2}+\varepsilon^{-1}_{3}\bar{\eta}_{3}\eta_{3}\right)}\frac{N!}{n_{1}!n_{2}!n_{3}!}\frac{\abs{\eta_{2}}^{2n_{2}}\abs{\eta_{3}}^{2n_{3}}}{\left(1+\abs{\eta_{2}}^{2}+\abs{\eta_{3}}^{2}\right)^{M}}.
\end{align*}
In the large $N$ limit, this expression simplifies to 
\begin{align*}
\bra{\bs n}\rho_{0}\ket{\bs n'} & \propto\delta_{ \bs n, \bs n'}\frac{e^{-\left(n_{2}+1\right)\log\left(\varepsilon^{-1}_{2}+1\right)-\left(n_{3}+1\right)\log\left(\varepsilon^{-1}_{3}+1\right)}}{4N}+O\left(\frac{1}{N}\right)^{2}.
\end{align*}
Thus the effective Hamiltonian of $\rho_{0} = e^{-H_\text{eff}}$ has a spectrum $e_{\bs n}= e_0 + n_{2} \log\left(\varepsilon^{-1}_{2}+1\right)+n_{3}\log\left(\varepsilon^{-1}_{3}+1\right)$,
with $e_0$ a normalization constant and $n_{2,3}=0,1,2...$ .

\section{Explicit Form of the $\text{IPR}_\phi$} \label{app:IPR}

In the number basis, $\ket{\boldsymbol n=\{n_{1},n_{2},n_{3}\}\in\Lambda_{N}}$,  
with $\Lambda_{N}=\left\{ \bs n:0\le n_{i=1,2,3}\le N\wedge\sum_{i=1}^{3}n_{i}=N\right\} $, integrals over phase-space in Eq.~(\ref{eq:IPR_phi}), can be evaluated explicitly using spherical coordinates in Eq.~(\ref{eq:inv_stereo}). This yields  
\begin{equation}
\int d\mu(\boldsymbol\eta) \, \mathbb{H}_\rho(\boldsymbol\eta) = \sum_{\bs n \in\Lambda_{N}}\bra{\bs n}\rho\ket{\bs n} = \tr{[\rho]},
\end{equation}
and
\begin{equation}
\int d\mu(\boldsymbol\eta)  \, \mathbb{H}_\rho(\boldsymbol\eta)^2 = \sum_{\bs n,\bs m,\bs o,\bs p\in\Lambda_{N}}\bra{\bs n}\rho\ket{\bs m}\bra{\bs o}\rho\ket{\bs p}C\left(\bs n,\bs m,\bs o,\bs p\right), 
\end{equation}
with 
\begin{equation}
C\left(\bs n,\bs m,\bs o,\bs p\right) =\delta_{\bs m+\bs o,\bs n+\bs p}\frac{(N+1)(N+2)(N!)^{2}}{(2N+2)!}\frac{\left(m_{1}+o_{1}\right)!\left(m_{2}+o_{2}\right)!\left(m_{3}+o_{3}\right)!}{\sqrt{m_{1}!m_{2}!m_{3}!n_{1}!n_{2}!n_{3}!o_{1}!o_{2}!o_{3}!p_{1}!p_{2}!p_{3}!}}. 
\end{equation}

\end{widetext}

\section{Leading Lyapunov exponent}
The leading Lyapunov exponent can be defined as  $$\lambda = \lim_{t\rightarrow \infty}  \frac{1}{t} \ln \frac{|\delta Y(t)|}{|\delta Y(0)|}, $$ where $$\delta Y(t) = M^t(Y_0) \delta Y(0)$$ is the time evolution of an initial infinitesimal deviation, $\delta Y(0)$, around the reference classical trajectory 
$$ Y_0(t) = \begin{pmatrix} \boldsymbol{\eta}(t)  \\ \boldsymbol{\bar{\eta}}(t) \end{pmatrix},  $$ and $M^t(Y_0)$ is the monodormy matrix respecting
$$ \partial_t \mathcal{M}^t(Y_0) = J^t(Y_0)  \mathcal{M}^t(Y_0), $$
with 
$$ J^{t}(Y_0) = \begin{pmatrix} \partial_{\boldsymbol{\eta}} \boldsymbol{X} &  \partial_{\boldsymbol{\bar{\eta}}} \boldsymbol{X}\\ \partial_{\boldsymbol{\eta}} \boldsymbol{\bar{X}} &  \partial_{\boldsymbol{\bar{\eta}}} \boldsymbol{\bar{X}} \end{pmatrix}, $$
the Jacobian of the classical evolution given by Eq.(\ref{semiclassical}), and initial condition $\mathcal{M}^{t=0}(Y_0)=\boldsymbol{1}$. 

Defining a normalized initial deviation vector, $\hat{n}=\frac{\delta \vec{Y}(0)}{| \delta \vec{Y}(0)|}$, we can  write 
\begin{equation}
     \lambda = \lim_{t\rightarrow \infty}  \frac{1}{t} \ln | \mathcal{M}^t(Y_0) \hat{n}|,
\end{equation}
which at large $t$, is dominated by the largest stability multiplier 
$\Lambda_1$ (for example $\Lambda_k$ for $k=1$), so the leading Lyapunov exponent is $\lambda =\lim_{t\rightarrow \infty}  \frac{1}{t} \ln | \Lambda_1|$ where $ \Lambda_1$  is the leading eigenvalue of the Jacobian matrix $J^t(Y_0) $.

\section{Liouvillian Gap Function}
\label{App-E-Liouv-Gap}

\begin{figure*}[th!]
\centering
\includegraphics[width=\linewidth]{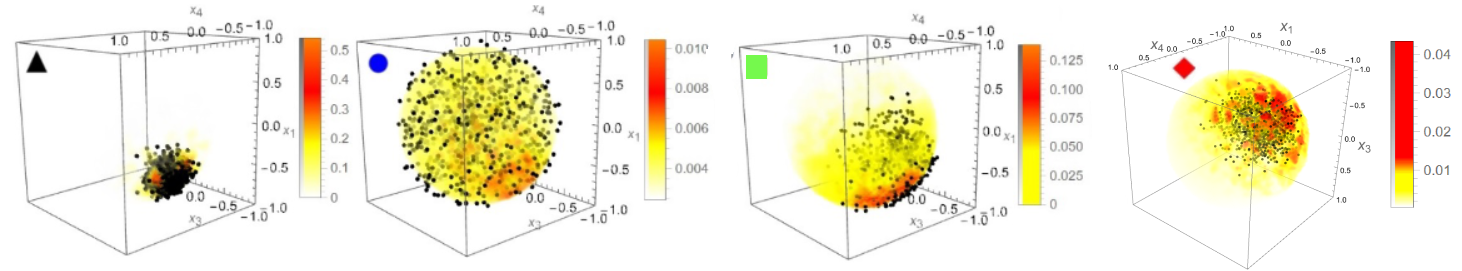} 
\caption{Husimi distribution and stochastic end point of trajectories. }
\label{fig:QuantumHsimiStocastic}
\end{figure*}

\subsection{Gapped system} To check the first part of the conjecture, we first considered values of two-particle interaction $U=1$ for which a rigid gap $\Delta=Re(\Lambda_{1})$ emerges in the Liouvillian spectrum ($\Lambda_{n}$) of the non-cyclic case, see  Fig.~\ref{fig:Gap}  (black triangle) a rigid gap as a function of size system. The importance of a gap in the first part of our conjecture resides in the fact that the lower eigenvalue of the Liouvillian superoperator occurs at the origin of the complex plane $\Lambda_{0} = 0$ and therefore the respective eigenvector defines one unique steady state and density matrix $\rho_0=\rho(t\rightarrow \infty)$. It means that all quantum trajectories must decay for this steady state with a decay relaxation time given by  $t_{r}\propto 1/\Delta$. In fact, for large times the density matrix behaves as $\rho(t)\propto \rho_0 + e^{-t/t_r}$. In other words, the gapped non-cyclic case is an interesting limit because the presence of the rigid gap in its spectrum prohibits the emergence of additional steady states when the size of the system goes to infinity.

\subsection{Gapless system} 
\begin{figure}[bh!]
\centering
\includegraphics[scale=0.35]{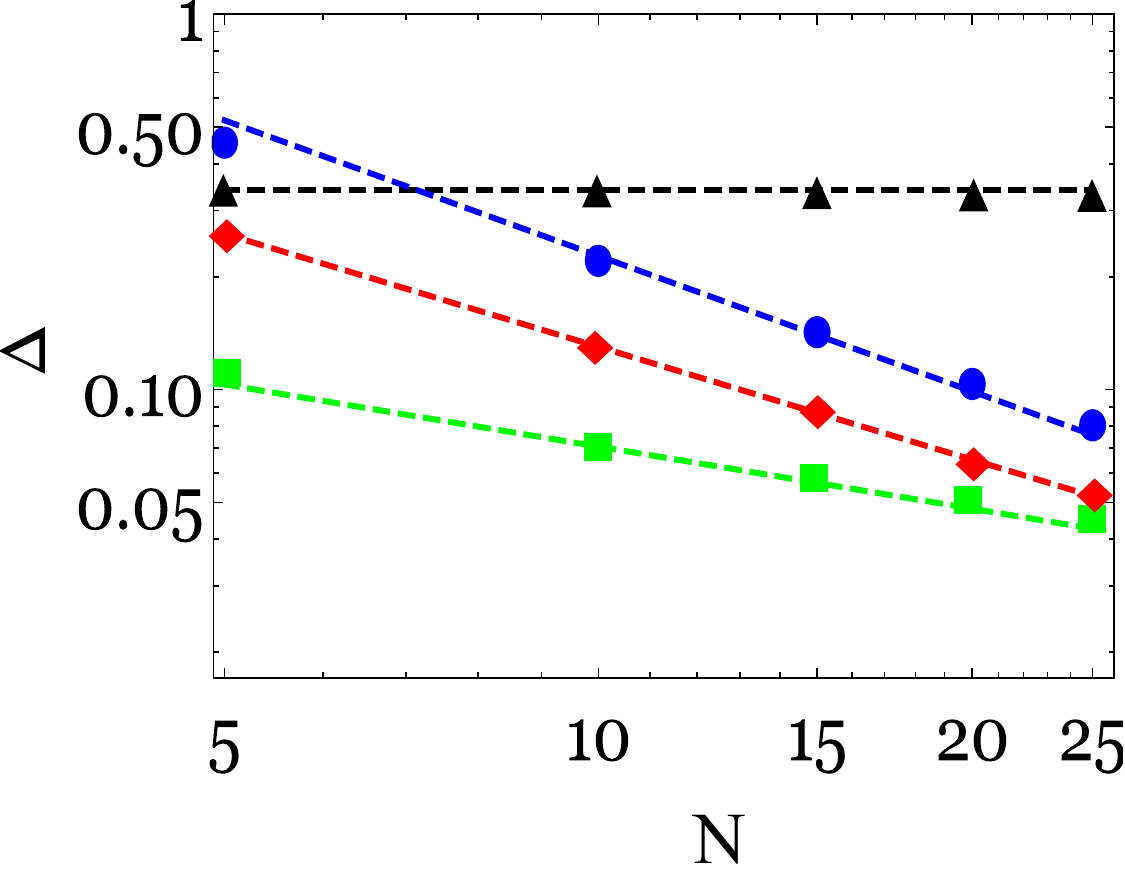} 
\caption{Gap as a function of system size $N$ for four cases of interest: the non-cyclic case (black triangle), and several cyclic cases (blue circle, green square, and red diamond). Only the non-cyclic case is gapped, while the cyclic ones are gapless.}
\label{fig:Gap}
\end{figure}

The second part of the conjecture states about the cyclic gapless case. 
When the gap is zero, the imaginary part of Liouvillian eigenvalues contributes to the emergence of quantum dynamics that never ends ($t_r \rightarrow \infty$). We show the gap as a function of systems size  $N$ for a set of cyclic cases in Fig.~\ref{fig:Gap} (blue circle, green square, and red diamond). The scale LogLog indicates a gap decay consistent with law $N^{-1}$ (blue, green, and red dashed lines).

\subsection{Liouvillian Spectrum} 

In Fig.~\ref{fig:Full-Liouv-Spec}, the Liouvillian spectrum is exhibited for each case: (a) non-cyclic $U=1$ (black color), cyclic $U=6$ (blue color) (b), cyclic $U=0.01$ (green color) (c) and cyclic $U=1$ (red color) (d). The top figure is the Liuvillian spectrum itself, and each middle panel represents a look more closely at the first  spectrum points for some values of the system size $N$. For the non-cyclic case, we can see that as $N$ is increased the spectrum points tend to converge to the fixed point (orange cross). These fixed points are obtained from the linearized Liouvillian, see Appendix~\ref{App-C-SC-DissD}. For the other cyclic cases, the effect of increasing $N$ moves the spectrum points to the right side of the spectrum, which reflects the gapless nature of the cyclic cases.

Studying the complex spacing ratio $z_{n}=\frac{\Lambda_{n}^{NN}-\Lambda_{n}}{\Lambda_{n}^{NNN}-\Lambda_{n}}$ of the full Liouvillian spectrum, we found a non-flat distribution for both non-cyclic and cyclic cases, see the bottom figures in Fig.~\ref{fig:Full-Liouv-Spec}. 
A non-flat anisotropic~\cite{Oganesyan2007,Atas2013,PhysRevX.Pedro} distribution of the Liouvillian eigenvalues is related to the quantum chaos emergence in the Liouvillian spectrum since it indicates a degree of level repulsion and angular dependence of the eigenvalue distribution~\cite{PhysRevX.Pedro}.  
In accordance with Fig.~\ref{fig:Full-Liouv-Spec}, both non-cyclic (gapped) and cyclic (gapless) cases exhibit correlated eigenvalues level statistics, and therefore its Liouvillian exhibits non-integrability features.
At this point, one interesting question arises; Does the steady state $\rho_{0}=\rho(t \rightarrow \infty )$ also exhibit or inherit the non-integrability characteristic of the Liouvillian spectrum? 
To answer this question, we compared the consecutive-level statistics of the density matrix with the distribution of the Lyapunov exponents obtained from the equations of motion in the semi-classical limit ($N \rightarrow \infty$). 

\begin{figure}[t!]
\centering
\includegraphics[width=9cm,height=4.0cm]{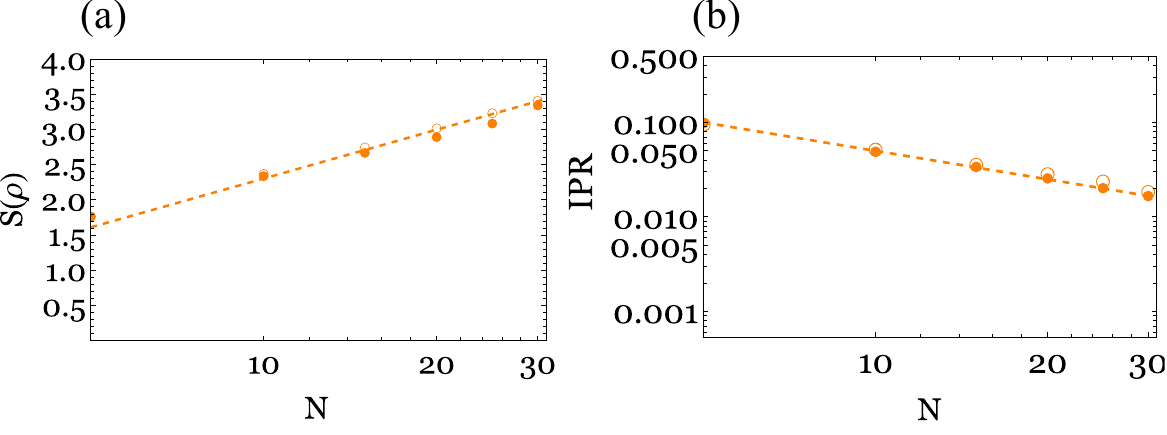}\label{FSa} 
\caption{ $SU(2)$ two wells model - Von Neumann entropy  $S(\rho)$  (a) and IPR (b) as a function of system size $N$.  }
\label{fig:SU2b}
\end{figure}

\section{Complex Spacing Level Statistics}
\label{App-F-Level-Statistics}

\begin{figure}[t!]
\centering
\includegraphics[width=8.5cm,height=4.0cm]{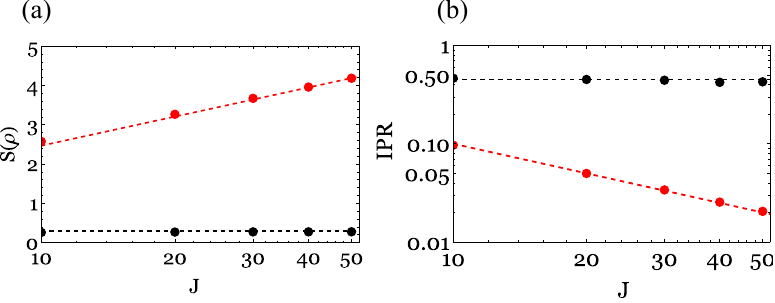}\label{FSa} 
\caption{$SU(2)$ - Von Neumann entropy  $S(\rho)$  (a) and IPR (b) as a function of system size $J$. The black and red points refer to the cases $h=0.25$ and $h=0.75$, respectively. Here,  $\gamma_{x}=\gamma_{y}=0$ and $\Gamma p =1$.  }
\label{fig:SU2}
\end{figure} 
%

The level repulsion  related to the complex spectrum can be obtained by the so-called complex spacing level ratio statistic $z_n$ defined in \cite{PhysRevX.Pedro} as $z_{n}=\frac{\Lambda_{n}^{NN}-\Lambda_{n}}{\Lambda_{n}^{NNN}-\Lambda_{n}}$. In Fig.~\ref{fig:Full-Liouv-Spec} we show the complex spacing ratio distribution on the complex plane for the non-cyclic and cyclic cases. For both cases, we find bitten donuts distribution. The donut distribution appears for a non-integrable Liouvilian superoperator.

\begin{figure*}
\centering
\includegraphics[width=1\linewidth,height=14cm]{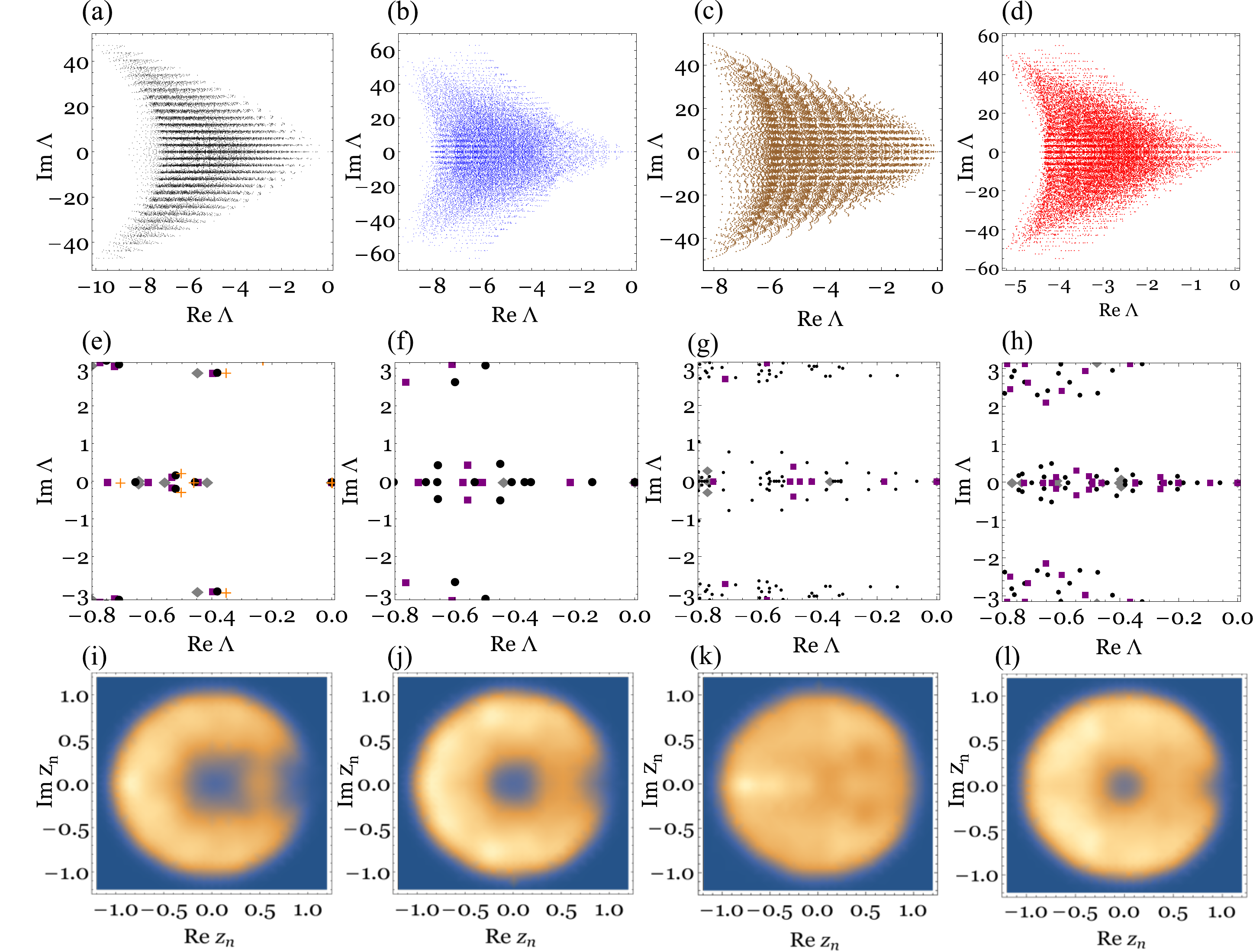}
\caption{Lindblad complex spectrum and complex spacing ratios $Z_n$. The first line shows the Lindblad spectrum behavior for attractor noncyclic case-I (a), full chaotic case-II (b), integrable $U=0$ case-V (c) and  mixed case-IV (d).  In the second line, we show some eigenvalues of the Lindblad superoperator that are close to the steady state for attractor noncyclic case-I (e), full chaotic case-II (f), integrable case-V with $U=0$ (g) and mixed case-IV (h). Here the colors are related to the $M$ value, $M=5$ (gray), $M=10$ (purple), $M=18$ (black). In special for the non-cyclic case-I, the orange crosses indicates the fixed points, see (e). The last column exhibits the density of complex spacing ratios for each case.
}
\label{fig:Full-Liouv-Spec}
\end{figure*}

\begin{figure*}
\centering
\includegraphics[width=14cm,height=14cm]{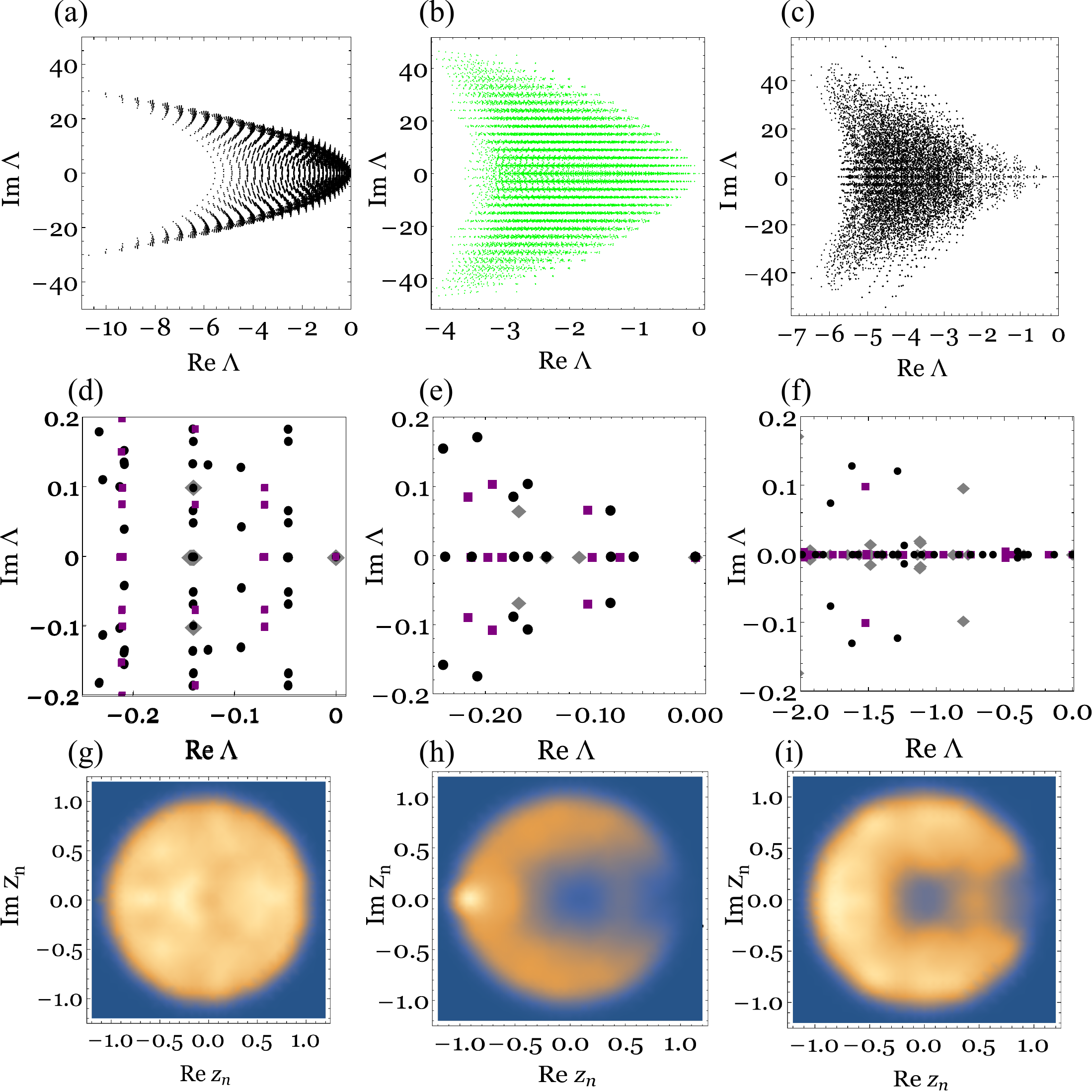}
\caption{Lindblad complex spectrum and complex spacing ratios $Z_n$. The first line shows the Lindblad spectrum behavior for $SU(2)$ case-VII (a), regular like case-III (b), full chaotic case-VI (c)  In the second line, we show some eigenvalues of the Lindblad superoperator that are close to the steady state for $SU(2)$ VII (d), integrable like case-III, full chaotic case-VI (e), The last column exhibits the density of complex spacing ratios for each case.
}
\label{fig:Full-Liouv-Spec}
\end{figure*}

\section{$SU(2)$ - Inverse Participation Ratio (IPR) and Von Neumann Entropy}
\label{App-I-IPR-VonNE}

In this section, we calculated the entropy and IPR in the $SU(2)$ limit. The $SU(2)$ limit is obtained after disconnecting well 1 from wells 2 and 3, by making $J_1=0,J_2=0$ and assuming that only the incoherent rates $w_{+3}$ and $w_{-3}$ are different from zero. In this limit the model possess only 2 wells connected and a $SU(2)$ algebraic structure emerges for the operators. In Fig.~\ref{fig:SU2b} we show the entropy and IPR$_\phi$ as a function of the number of bosons for $U=1$. Due to the presence of a strong symmetry which induces a constant of motion, the entropy increases linearly with $ln(N)$ and the IPR decreases as $\frac{1}{N}$. In the case of IPR, the strong symmetry is related to a presence of constant of motion which restrict the region in phase-space where the trajectories can evolve.

We also calculated the IPR and entropy for the $SU(2)$ Lipkin-Meshkov-Glick (LMG) model with Markovian dissipation.  
The (LMG) model presents a well-known integrable steady state (e.g., a single steady state and periodic trajectories)~\cite{Ribeiro2019}, therefore the IPR and entropy obtained in this section can be compared with the results of the $SU(3)$ model in the limit of integrable steady states.  

The Hamiltonian and Liovillian superoperator can be written as 
\begin{equation}\label{Hsu2}
    H =-h \cdot S+\frac{1}{2s}(\gamma_x S^2_x+\gamma_y S^2_y)
\end{equation}
and 
\begin{equation}\label{LiouSu2}
    \mathcal{L}=-i[H,\rho]+\sum_i(W_i\rho W^\dagger_i-\frac{1}{2} \left\{ \rho,W_i W_i \right\} ).
\end{equation}

We considered only the case where the quantization axis of the operator is taken along the polarization of the tip, and for the parallel case. The spin operators $S_{\eta=x,y,z}$ obey the $SU(2)$ commutation relations with $S \cdot S=s(s+1)$. In Eq.~\ref{Hsu2} and \ref{LiouSu2} $h$ is the magnetic field, and the jump operators are $W_z=\sqrt{\frac{\Gamma}{2s}}S_z$, $W_+=\sqrt{\frac{\Gamma(1-p)}{4s}}S_+$, $W_-=\sqrt{\frac{\Gamma(1+p)}{4s}}S_-$, where $\Gamma$ is the rate of the quantum jumps.

As showed in Ref.~\cite{Ribeiro2019}, for $\gamma_{x}=\gamma_{y}=0$, $\Gamma p =1$ and $h=0.25$, we obtain one phase with single steady state. On the other hand, for $\gamma_{x}=\gamma_{y}=0$, $\Gamma p =1$ and $h=0.75$, we find a different phase with periodic trajectories in phase-space. 
The entropy and IPR as a function of $J$ for each phase can be visualized in Fig.~\ref{fig:SU2}-(a) and (b) for $h=0.25$ (black points) and $h=0.75$ (red points), respectively. As we can see, the entropy as a function of $J$ is constant when the phase exhibits trajectories that fall to a steady state (black points-(a)), while for a phase in the presence of periodic trajectories $h=0.75$  (red points-(b)), the entropy increases linearly with $Log s$. The IPR as a function of $s$ also exhibits a constant behaviour for $h=0.25$  (black points-(b)), while for $h=0.75$ (red points-(b)) the IPR decays following the law $1/J$.

\clearpage
\nocite{*}
\bibliographystyle{unsrt}
\bibliography{biblio2}{}


\end{document}